\documentclass[final,1p,times,twocolumn]{elsarticle}

\usepackage{amssymb}
\usepackage{amsfonts}
\usepackage{graphicx}
\usepackage{amsmath}

\journal{Computer Physics Communications}

\begin{document}

\begin{frontmatter}



\title{Self-consistent GW method: O(N) algorithm for polarizability and self energy}

\author{A.~L.~Kutepov$^{a}$\corref{a}}

\cortext[a] {Corresponding author.\\\textit{E-mail address:} akutepov@bnl.gov}
\address{$^{a}$Condensed Matter Physics and Materials Science Department, Brookhaven National Laboratory, Upton, NY 11973}

\begin{abstract}
An efficient implementation of the self-consistent GW method in the FlapwMBPT code \cite{flapwmbpt} is presented. It features the evaluation of polarizability and self-energy which scales linearly with respect to the system size. Altogether the computational time scaling was measured to be between linear and quadratic in the applications to silicon supercells with up to 72 atoms. Application to such materials as paracostibite CoSbS, supercells of La$_{2}$CuO$_{4}$ (up to 56 atoms) and SmB$_{6}$, illustrate the potential of the approach in computational material science.

\end{abstract}


\end{frontmatter}



\section{Introduction}
\label{intr}

The so called GW method was introduced by Lars Hedin in his seminal paper \cite{pr_139_A796} more than 50 years ago. Its non-self-consistent variant, which often is abbreviated as $G_{0}W_{0}$ is probably the most popular method for the evaluation of the electronic structure in weakly and moderately correlated materials. It has been implemented in many computer codes as an efficient perturbative correction to the electronic structure usually calculated at the DFT (Density Functional Theory) level. Since its first pseudopotential-based implementation and applications to simple materials \cite{prb_25_2867,prb_34_5390}, considerable efforts were undertaken to enhance the computational efficiency of the method \cite{prb_78_113303,jcp_133_164109,prb_79_245106,prl_100_147601,prb_81_115104,prb_81_115105,prb_78_085125,
prb_82_041103,jcp_135_074105,prb_94_165109,prl_113_076402,arx.1904.10512,sr_6_36849}. Also, the $G_{0}W_{0}$ approach was adapted to all-electron basis sets such as FLAPW (Full-potential Linearized Augmented Wave) method \cite{prb_93_115203,prb_74_045104,prb_83_081101,prb_84_039906,prb_94_035118,prb_80_041103}.

Despite its popularity and sufficiently low computational cost, the $G_{0}W_{0}$ approach is seriously limited when one needs a real prediction of the basic features of electronic structure. Different starting points (such as LDA, GGA, Hybrid functionals, LDA+U) are required depending on a particular material and it is hard to make a decision on the starting point without having some experimental input available. For instance, using Perdew-Burke-Ernzerhof (PBE) functional \cite{prl_77_3865} as a starting point for the $G_{0}W_{0}$ calculations of the band gap in simple covalent semiconductors and insulators might be considered as a sufficiently successful strategy. The combination PBE+G$_{0}$W$_{0}$ provides the gaps with typical underestimation (as compared to the experimental gaps) within 10-15\% \cite{prb_93_115203}. Application of $G_{0}W_{0}$ to d- and f-electron oxides requires the usage of different approximation (LDA+U, \cite{prb_44_943}) as a starting point \cite{prb_97_245132}. LDA+U method, in its own turn, has an adjustable parameter (Hubbard U) which complicates its usage in the combination LDA+U+$G_{0}W_{0}$ even further. Other starting points are also possible. For example, the band gap in the widely studied insulating monoclinic phase of vanadium dioxide VO$_{2}$ can be reproduced with the COHSEX approach \cite{pr_139_A796} as a starting point, whereas the $G_{0}W_{0}$ based on the LDA doesn't open the gap \cite{prl_99_266402}. Obviously, the development of efficient self-consistent GW-based approaches which eliminate the dependence on the starting point is important.

Several self-consistent implementations of the GW method (scGW) are now available \cite{flapwmbpt,ecalj,questaal,vasp}. Their applications to real materials are, however, limited by considerable computational cost and memory requirements. Normally one applies them to materials with less than 10-12 atoms per unit cell. At the same time, a surge in the experimental exploration of new materials which we are witnessing nowadays deals with 20-30 (and more) atoms per unit cell. Thus, an extension of the applicability of scGW (or its highly successful quasiparticle version QSGW \cite{prb_76_165106}) to the solids with large unit cells is mandatory.

High computational cost of the self-consistent GW method originates mostly from the fact that many existing GW codes (both the $G_{0}W_{0}$ and scGW) still use a naive implementation of polarizability P and self-energy $\Sigma$. This implementation uses frequencies and momenta as basic variables and scales as $O(N^{4}N_{\omega}^{2}N_{k}^{2})$ (N being the system size, $N_{\omega}$ is the number of frequencies, and $N_{k}$ stands for the number of points in the Brillouin zone (momenta)).

Certain progress was achieved during the last 20 years. First, the evaluation of P and $\Sigma$ is more efficient when one uses imaginary time instead of frequency (scales linearly with respect to the number of time points) \cite{prl_74_1827,prb_80_041103}. Secondly, the evaluation of P and $\Sigma$ benefits a lot when the real space is used instead of the \textbf{k}-space. The real space implementation scales formally as $O(N^{2}N_{k})$. The first implementation of it was announced more than 20 years ago \cite{prl_74_1827} in the context of the plane wave basis set and pseudopotential approximation. The first implementation of the real space technique for all-electron method and MT geometry (FLAPW basis set) was done in the code FlapwMBPT \cite{flapwmbpt} and published a few  years ago \cite{prb_85_155129}. Recently, a similar technique was also included in the code VASP \cite{vasp} in the context of the PAW (Projected Augmented Wave) method \cite{prb_94_165109}.

In this work, the advantages of using the real space in the evaluation of P and $\Sigma$ are explored further. Namely, not only the linear scaling with respect to the system size is introduced, but also it is shown that the real space provides greater flexibility for the choice of the basis set which makes calculations faster and dramatically alleviates the memory requirements.

\section{Methodology}
\label{GW}

Let us begin the presentation of the GW implementation with a brief account of the self-consistent GW method in its classic form which is written in momenta+frequency representation. Each iteration of the GW method takes some approximate Green's function G as an input and proceeds with the following steps. First one evaluates the polarizability

\begin{align}  \label{P_old}
P_{IJ}^{\mathbf{q}}(\nu)= \sum_{\alpha}\sum_{\mathbf{k}\omega} \sum_{\lambda\lambda''}
\langle M_{I}^{\mathbf{q}}|\Psi_{\lambda}^{\alpha\mathbf{k}}\Psi_{\lambda''}^{\alpha\mathbf{k-q}}\rangle  \sum_{\lambda'}G_{\lambda\lambda'}^{\alpha\mathbf{k}}(\omega )\sum_{\lambda'''}
G_{\lambda''\lambda'''}^{\alpha;\mathbf{k-q}}(\omega -\nu)
\langle \Psi_{\lambda'}^{\alpha\mathbf{k}}\Psi_{\lambda'''}^{\alpha\mathbf{k-q}}|M_{J}^{\mathbf{q}}\rangle ,
\end{align}
where $\alpha$ stands for spin, $\mathbf{k}$ and $\mathbf{q}$ are momenta (points in the Brillouin zone), $\nu$ and $\omega$ are bosonic and fermionic Matsubara frequencies correspondingly, $\lambda,\lambda',\lambda'',\lambda'''$ stand for band indexes. $I$ and $J$ are product basis indexes, $M$ are product basis functions and $\Psi$ are the band Bloch's functions. The expression (\ref{P_old}) scales as $O(N^{4})$ with respect to the system size $N$, as $O(N^{2}_{\nu})$ with the number of frequencies, and as $O(N^{2}_{\mathbf{k}})$ with the number of momenta. It is very time consuming in both limits of the very large systems (small $N_{\mathbf{k}}$ but large $N$) and of the small systems (small $N$ but large $N_{\mathbf{k}}$).

Next step in the self-consistent GW method is the evaluation of the screened interaction $W$:

\begin{align}  \label{W_1}
W^{\mathbf{q}}_{IJ}(\nu)=V^{\mathbf{q}}_{IJ}+\sum_{KL} V^{\mathbf{q}}_{IK}P^{\mathbf{q}}_{KL}(\nu)
W^{\mathbf{q}}_{LJ}(\nu).
\end{align}

This expression scales linearly with respect to the number of momenta and frequencies, and its scaling is cubic with respect to the system size. Quantities $V^{\mathbf{q}}_{IJ}$ in (\ref{W_1}) represent matrix elements of the bare Coulomb interaction.

The third step is the evaluation of self energy:

\begin{align}  \label{S_old}
\Sigma_{\lambda\lambda'}^{\alpha\mathbf{k}}(\omega)=-\sum_{\mathbf{q}}\sum_{\lambda''\lambda'''}\sum_{IJ}
\langle \Psi_{\lambda}^{\alpha\mathbf{k}}|\Psi_{\lambda''}^{\alpha\mathbf{k-q}}M_{I}^{\mathbf{q}}\rangle G_{\lambda''\lambda'''}^{\alpha\mathbf{k-q}}(\omega-\nu)W_{IJ}^{\mathbf{q}}(\nu)
\langle \Psi_{\lambda'}^{\alpha\mathbf{k}'}|\Psi_{\lambda'''}^{\alpha\mathbf{k-qt}}M_{J}^{\mathbf{q}}\rangle,
\end{align}
which scales similarly to the expression (\ref{P_old}) for polarizability.

Finally one has to solve the Dyson equation for Green's function

\begin{align}  \label{G_1}
G^{\alpha\mathbf{k}}_{\lambda\lambda'}(\omega)=G^{0,\alpha\mathbf{k}}_{\lambda\lambda'}(\omega)+\sum_{\lambda''\lambda'''} G^{0,\alpha\mathbf{k}}_{\lambda\lambda''}(\omega)\Sigma^{\alpha\mathbf{k}}_{\lambda''\lambda'''}(\omega)G^{\alpha\mathbf{k}}_{\lambda'''\lambda'}(\omega),
\end{align}
which formally scales similarly to the expression (\ref{W_1}) for the screened interaction, but in practice is less time consuming because usually the number of band states is a few times smaller than the size of the product basis. Quantities $G^{0,\alpha\mathbf{k}}_{\lambda\lambda''}(\omega)$ in (\ref{G_1}) are the components of the Green function in the Hartree approximation.

From the comparison of scalings of the above four equations one can easily conclude that in order to make calculations faster one needs to optimize the evaluation of P and $\Sigma$. As it was already mentioned in the Introduction, the first serious step in that direction was undertaken by Rojas et al. \cite{prl_74_1827} who implemented the GW method in the real space + imaginary time representation with the pseudopotential approximation and plane wave basis set. The scaling of the evaluation of the polarizability and the self energy was improved to $O(N^{2}N_{\omega}N_{k})$ with obvious advantages. In our previous paper \cite{prb_85_155129} (the corresponding algorithm will be called as the old version here), we implemented the self-consistent GW approach for all electron full potential method (FLAPW+LO basis set) in the real space + imaginary (Matsubara) time representation which resulted in enormous saving of computer time in the applications to realistic materials. Because the basic formulae of the implementation in the Ref. \cite{prb_85_155129} are very similar to the ones used in the new methodology which is presented in this work it is convenient to give them here (formulae (\ref{P_MM})-(\ref{S_II}) below). Muffin-tin geometry associated with the LAPW method breaks the expressions for P and $\Sigma$ into three distinctive cases corresponding to where the two space arguments belong (MT sphere or interstitial (INT)). In the formulas given below, the following notations are used: $\mathbf{t},\mathbf{t}'$ for the coordinates of the centers of the MT spheres; $K,K'$ for the product basis set indexes inside the muffin-tin spheres; $\varphi_{L}^{\alpha\mathbf{t}}$ for the one-electron orbitals inside the MT spheres with $L,L',L'',L'''$ representing all quantum numbers to distinguish them; $\tau$ for the Matsubara (imaginary) time defined in the range [0:$\beta$]; $\mathbf{r},\mathbf{r}'$ for the points on the real space grid in the interstitial region; $\widetilde{W}$ is the screening part of the interaction $W$ ($W=V+\widetilde{W}$). Formulae (\ref{S_MM})-(\ref{S_II}) represent the evaluation of the dynamic (frequency dependent) part of self energy. The static (exchange) part is evaluated according to the same expressions but with the replacement $\widetilde{W}(\tau)\rightarrow V$ and $G(\tau)\rightarrow -G(\beta)$.

Polarizability (MT-MT):
\begin{align}  \label{P_MM}
P_{\mathbf{t}K;\mathbf{t}'K'}^{\mathbf{R}}(\tau)=-\sum_{\alpha} \sum_{LL''}\langle M_{K}^{\mathbf{t}}|\varphi_{L}^{\alpha\mathbf{t}}\varphi
_{L''}^{\alpha\mathbf{t}}\rangle \sum_{L'}G_{\mathbf{t}L;\mathbf{t}'L'}^{\alpha\mathbf{R}}(\tau )\sum_{L'''}
G_{\mathbf{t}L'';\mathbf{t}'L'''}^{\alpha;\mathbf{R}}(\beta -\tau)  \langle \varphi_{L'}^{\alpha\mathbf{t}'}\varphi_{L'''}^{\alpha\mathbf{t}'}|M_{K'}^{\mathbf{t}'}\rangle ,
\end{align}
(MT-INT):
\begin{align}  \label{P_MI}
P^{\mathbf{R}}_{\mathbf{t}K;\mathbf{r}'}(\tau)=-\sum_{\alpha}\sum_{LL'} \langle
M^{\mathbf{t}}_{K}|\varphi^{\alpha\mathbf{t}}_{L}\varphi^{\mathbf{t}}_{L'}\rangle
G^{\alpha\mathbf{R}}_{\mathbf{t}L;\mathbf{r}'}(\tau)G^{\alpha\mathbf{R}}_{\mathbf{t}L';\mathbf{r}'}(\beta-\tau),
\end{align}
(INT-INT):
\begin{align}  \label{P_II}
P^{\mathbf{R}}_{\mathbf{r}\mathbf{r}'}(\tau)=-\sum_{\alpha}G^{\alpha\mathbf{R}}_{\mathbf{r}\mathbf{r}'}(\tau)
G^{\alpha\mathbf{R}}_{\mathbf{r}\mathbf{r}'}(\beta-\tau).
\end{align}

Self energy (MT-MT):
\begin{align}  \label{S_MM}
\Sigma_{\mathbf{t}L;\mathbf{t}'L'}^{\alpha\mathbf{R}}(\tau)=-\sum_{L''L'''}\sum_{KK'}\langle \varphi_{L}^{\alpha\mathbf{t}}|\varphi
_{L''}^{\alpha\mathbf{t}}M_{K}^{\mathbf{t}}\rangle G_{\mathbf{t}L;\mathbf{t}'L'}^{\alpha\mathbf{R}}(\tau )
\widetilde{W}_{\mathbf{t}K;\mathbf{t}'K'}^{\mathbf{R}}(\beta-\tau)   \langle \varphi_{L'}^{\alpha\mathbf{t}'}|\varphi_{L'''}^{\alpha\mathbf{t}'}M_{K'}^{\mathbf{t}'}\rangle ,
\end{align}
(MT-INT:
\begin{align}  \label{S_MI}
\Sigma^{\alpha\mathbf{R}}_{\mathbf{t}L;\mathbf{r}'}(\tau)=-\sum_{L'K} \langle
\varphi^{\alpha\mathbf{t}}_{L}|\varphi^{\mathbf{t}}_{L'}M^{\mathbf{t}}_{K}\rangle
G^{\alpha\mathbf{R}}_{\mathbf{t}L;\mathbf{r}'}(\tau)\widetilde{W}^{\mathbf{R}}_{\mathbf{t}K;\mathbf{r}'}(\beta-\tau),
\end{align}
(INT-INT):
\begin{equation}  \label{S_II}
\Sigma^{\alpha\mathbf{R}}_{\mathbf{r}\mathbf{r}'}(\tau)=-G^{\alpha\mathbf{R}}_{\mathbf{r}\mathbf{r}'}(\tau )
\widetilde{W}^{\mathbf{R}}_{\mathbf{r}\mathbf{r}'}(\beta -\tau ).
\end{equation}

Quadratic scaling of the above formulas follows from the fact that the computations in each of them are proportional to the number of pairs $(\mathbf{t},\mathbf{t}')$, $(\mathbf{t},\mathbf{r}')$, or $(\mathbf{r},\mathbf{r}')$ which grows quadratically when the unit cell increases. Linear scaling with respect to the number of time points and of the $\mathbf{k}$-points (which is equal to the number of $\mathbf{R}$-vectors in the old version) is also obvious. The implementation of the self consistent GW method in the real space + Matsubara time representation allowed considerable savings of computer time. However, the need to store quantities like $F^{\mathbf{k}}_{\mathbf{r}\mathbf{r}'}(\tau)$ (or like $F^{\mathbf{k}}_{\mathbf{t}L\mathbf{t}'L'}(\tau)$) where $F$ stands for any two-point quantity (G, P, $\Sigma$, or W) was limiting the applications to the unit cells with no more than 10-15 atoms. The principal reason for the exceedingly large amount of information for storing is that the approximations (such as the size of the basis set) were introduced in the reciprocal space which didn't provide a flexibility in the choice of a basis set size. The basis set size was essentially the same for all $\mathbf{k}$-points. Real space, which formally provides such flexibility (see below) was used in the old version only as an intermediate step in order to accelerate the calculations.

\begin{figure}[t]
\centering
\includegraphics[width=9.0 cm]{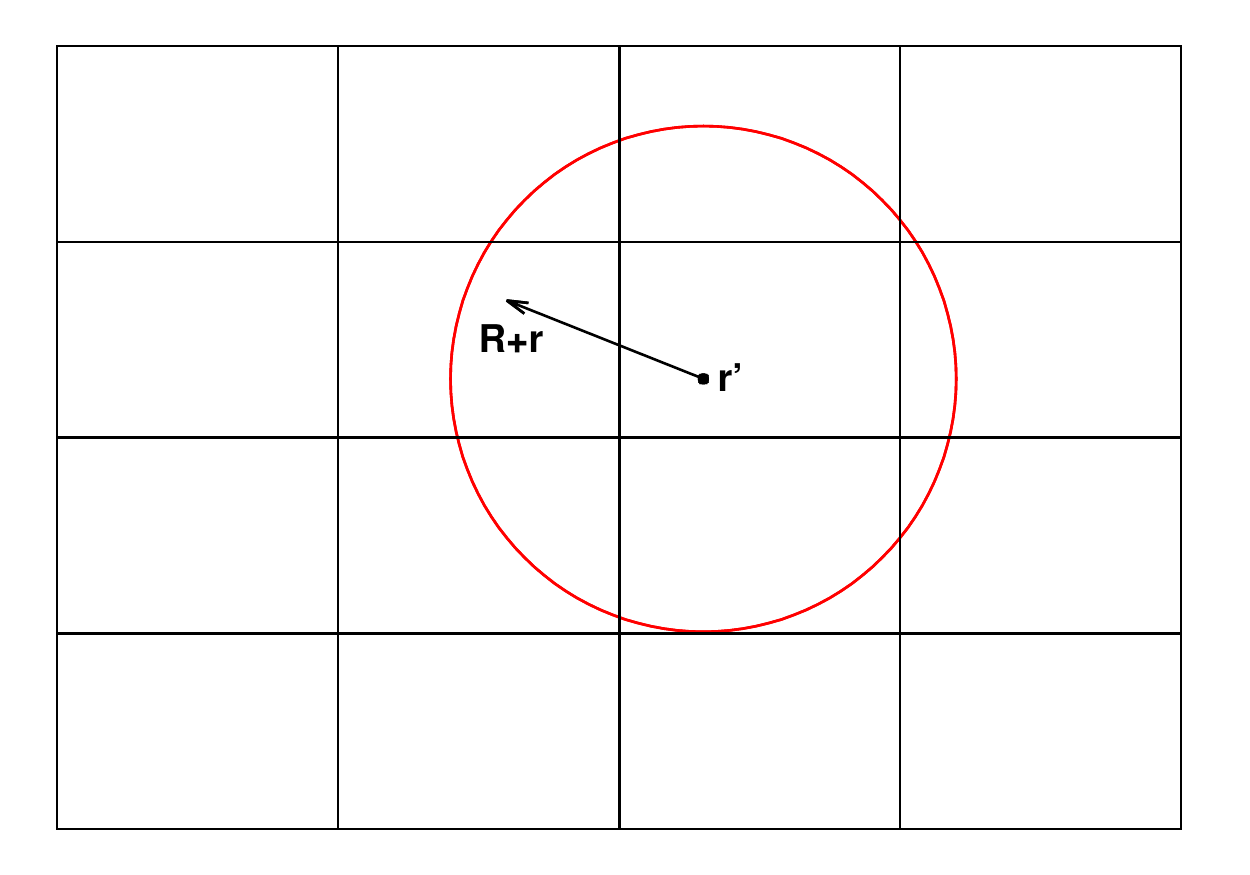}
\caption{Graph illustrating the basic approximation. Vertical and horizontal lines correspond to the boundaries of unit cells. For any given point $\mathbf{r}'$ (center of the sphere in the figure) the two point functions $F_{\mathbf{r},\mathbf{r}'}$ are nonzero only for $\mathbf{r}$ inside the sphere of a certain radius (red circle). The number of unit cells which the limiting sphere encloses or intersects in each direction defines the mesh of $\mathbf{k}$-points in the Brillouin zone. See the text for more details.} \label{basic_appr}
\end{figure}

In this work it is proposed to use real space as a principal representation, where all important cutoffs are introduced. The quantities for storing now look like $F^{\mathbf{R}}_{\mathbf{r}\mathbf{r}'}(\tau)\equiv F_{\mathbf{R}+\mathbf{r}\mathbf{r}'}(\tau)$, where the real space points $\mathbf{r}$ and $\mathbf{r}'$ belong to the unit cells (generally different) and the translation vector $\mathbf{R}$ measures the distance between the two unit cells. The essential difference of the new approach as compared to the old version of the code FlapwMBPT consists in the fact that the list of $\mathbf{R}$-vectors is defined not from the given $\mathbf{k}$-mesh but from the following consideration which naturally leads to the linear scaling with respect to the system size. Namely, one accepts an approximation $F^{\mathbf{R}}_{\mathbf{r}\mathbf{r}'}(\tau)=0$ for all pairs $\mathbf{R+r},\mathbf{r}'$ with $|\mathbf{R}+\mathbf{r}-\mathbf{r}'|>R_{max}$. $R_{\max}$ here is a parameter eventually defining the density of the $\mathbf{k}$-points in the Brillouin zone. Now the requirement for the number of $\mathbf{k}$-points is that it should be the same or larger than the number of unit cells which are inside of or cross the sphere $R_{max}$. This ensures the absence of periodic images inside the sphere. For example, in Figure \ref{basic_appr} the minimal (two-dimensional in this case) $\mathbf{k}$-mesh should be at least $3\times 3$ as the sphere $R_{max}$ encloses or crosses three unit cells in each direction. Formally, the convergence is achieved when function at $R_{max}$ is very small. By making $R_{max}$ independent on the system size (which is physically sound) the scaling becomes $O(mN)$ where $m$ is the number of $\mathbf{R}+\mathbf{r}$ points inside sphere $R_{max}$. The number $m$ is independent on the system size and, correspondingly, the approach has linear scaling.

The principal advantage of the new methodology comes however not from its linear scaling, but from the flexibility in the selection of the basis set size which it provides. The idea is that the change in any two-point function $F^{\mathbf{R}}_{\mathbf{r}\mathbf{r}'}$ when the vectors $\mathbf{R+r}$ and/or $\mathbf{r}'$ change is much larger when $\mathbf{R+r}$ and $\mathbf{r}'$ are close to each other than when they are distant. In practice, it means that the density of $\mathbf{r}$ and $\mathbf{r}'$ points can be considerably reduced when they belong to the distant unit cells. In the LAPW method, this statement is directly relevant to the interstitial region. Inside the MT spheres, the functions are expanded in spherical harmonics. Again the cutoff $L_{max}$ for this expansion can be considerably reduced when two MT spheres are distant.

\begin{figure}[t]
\centering
\includegraphics[angle=270,width=9.0 cm]{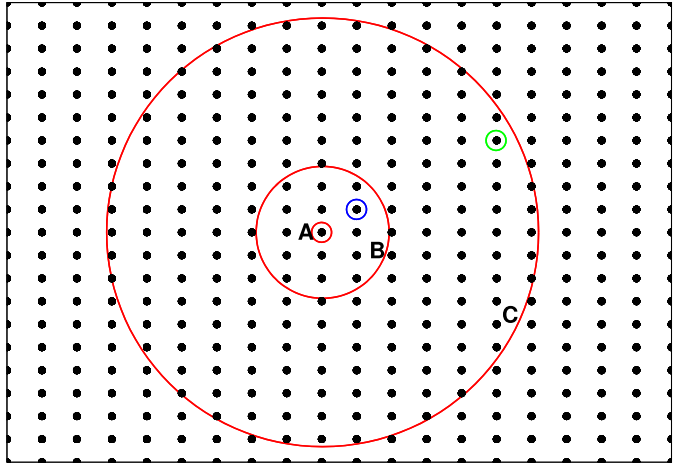}
\caption{Illustration of the flexibility of the real space for a basis set selection. Dots on the grid correspond to the centers of atoms. When two space arguments of a two point function 
$F_{\mathbf{r},\mathbf{r}'}$ belong to the same central sphere (sphere A in the figure) the function is expanded in a largest basis set. When one space argument belongs to sphere A but another one belongs to sphere B (for instance blue circle) the basis set can be reduced. Finally, when one space argument belongs to sphere A but another one belongs to sphere C (for instance green circle) the basis set is the smallest.} \label{spheres1}
\end{figure}

The present implementation of the above ideas in the code FlapwMBPT uses three levels of approximation related to the basis sets inside muffin spheres. Figure \ref{spheres1} illustrates it. The largest basis set is used when both space arguments belong to the same MT sphere (small sphere A in the middle of the figure). The intermediate size of the basis set is used when the distance between the centers of the MT spheres doesn't exceed the radius of sphere B in the figure. For example, this situation is realized when argument $\mathbf{r}'$ belongs to the central (red) MT sphere A but argument $\mathbf{r}$ belongs to the blue sphere which is one of the MT spheres inside limiting sphere B. Finally, the smallest basis set is used when the distance between the muffin spheres is larger than the radius of sphere B but smaller than the radius of sphere C. An example would be the pair consisting from the red MT sphere at the center and the green MT sphere inside limiting sphere C. When space argument belongs to the interstitial region, however, there is obviously no central A sphere. In this case only spheres B and C are used when one constructs the grids of $\mathbf{r}$ and $\mathbf{r}'$ points.

One iteration of the new implementation formally proceeds in precisely the same way as it is in the older version of the FlapwMBPT code. It was described in details in the Ref. \cite{prb_85_155129} (fully relativistic variant) and in the Ref. \cite{cpc_219_407} (non-relativistic variant). Briefly one follows the formulae (\ref{P_old})-(\ref{G_1}). However, the equation (\ref{P_old}) is replaced with efficient (order $O(N)$ implementation of equations (\ref{P_MM})-(\ref{P_II}) and the equation (\ref{S_old}) is replaced with the corresponding efficient implementation of the equations (\ref{S_MM})-(\ref{S_II}). For the evaluation of W (Eqn. \ref{W_1}) and of Green's function (Eqn. \ref{G_1}) the LAPACK or SCALAPACK libraries are used. One more thing needs to be mentioned here. The transformations between $\mathbf{R}$ and $\mathbf{k}$ spaces are also performed more efficiently in the new implementation as compared to the old version. This is because the complexity of these transformations is proportional to the number of pairs ($\mathbf{R+r},\mathbf{r}'$ for instance) which grows as $O(N)$ in the new algorithm versus $O(N^{2})$ in the old one. Also, which is even more important for the efficiency, the basis set size inside the muffin tin spheres (or the density of grid points in the interstitial region) is more compact in the new implementation.

\section{Convergence tests}
\label{conv}

In this section, a brief account of the convergence of the results with respect to the most important setup parameters of the new implementation is given. They are the radii of spheres B and C (see Fig. \ref{spheres1}) and the spherical harmonics cutoff $L_{max}$ for the product basis expansions (both for atoms in sphere B and in sphere C). The possibility of a reduction in the density of the \textbf{r}-points in the interstitial region for remoted unit cells is similar to the possibility of a reduction in the $L_{max}$ and, correspondingly, will not be presented.

\begin{figure}[h]       
    \fbox{\includegraphics[width=6.0 cm]{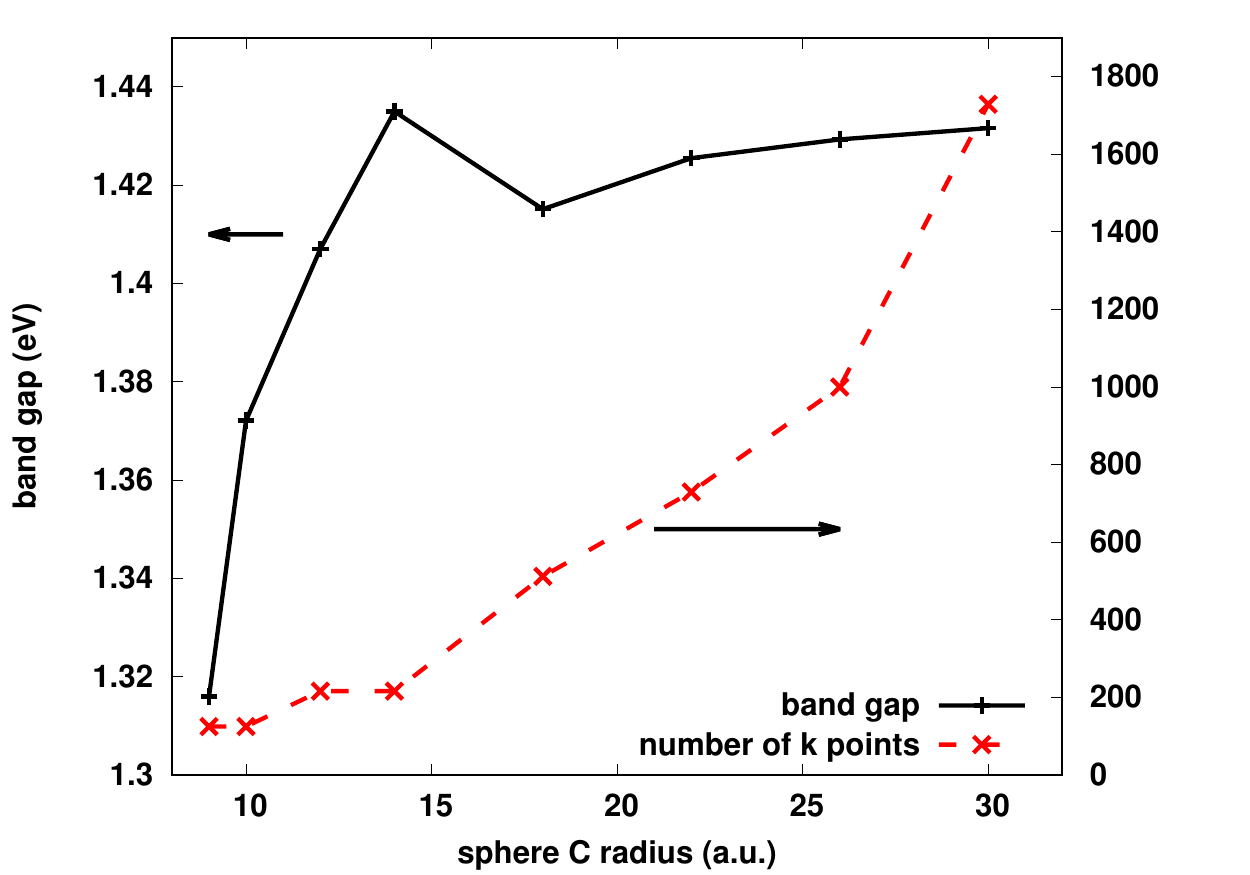}}   
    \hspace{0.02 cm}
    \fbox{\includegraphics[width=6.0 cm]{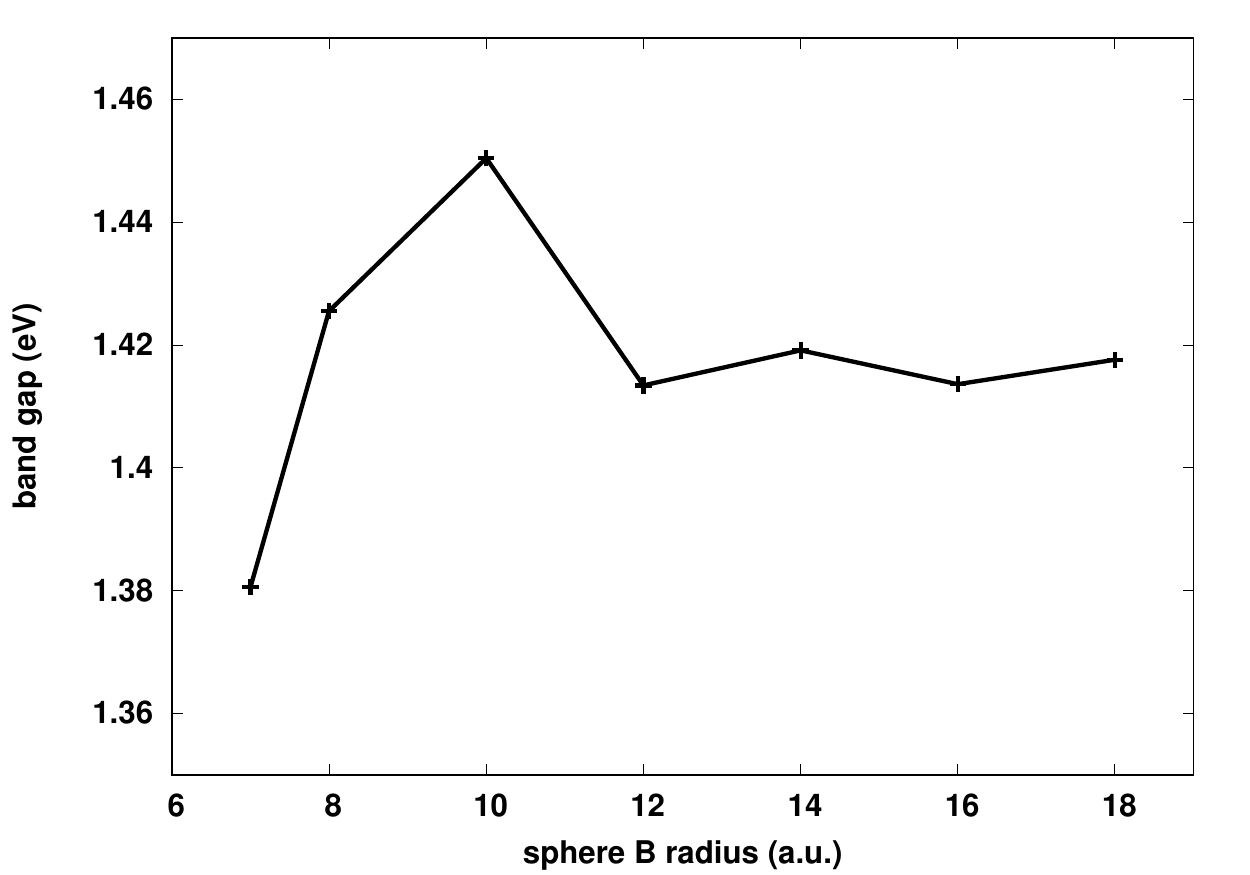}}
    \caption{Left: Convergence of the calculated band gap of Si with respect to the radius of sphere C (see Fig. \ref{spheres1}). The radius of sphere B was fixed at 8 a.u.. Right: Convergence of the calculated band gap of Si with respect to the radius of sphere B (see Fig. \ref{spheres1}). The radius of sphere C was fixed at 24 a.u.}
    \label{conv_rbc}
\end{figure}

The left part of figure \ref{conv_rbc} represents the calculated band gap of silicon as a function of the radius of sphere C. The radius of sphere B was fixed at 8 a.u. The cutoff $L_{max}$ was fixed at 6 (spheres A and B) and at 2 (outside of sphere B but inside of sphere C). Other setup parameters, such as the number of functions in the muffin tin spheres and the number of plane waves in the interstitial region were also fixed at the level ensuring that an uncertainty of the calculated band gap was no more than 0.02-0.03 eV. As it was explained earlier, an increasing of the radius of sphere C automatically changes the density of $\mathbf{k}$-points in the Brillouin zone. This is reflected in the Fig.\ref{conv_rbc} as well. As one can see, the change in the number of $\mathbf{k}$-points is not necessary monotonic: it happens only when sphere C crosses additional unit cells (see Fig. \ref{basic_appr}). The band gap converges within 0.02eV when sphere C radius reaches 14 a.u. It corresponds to 216 $\mathbf{k}$-points ($6\times 6\times 6$ mesh in the Brillouin zone) which reflects usual rate of convergence for this material.

The right part of figure \ref{conv_rbc} shows the band gap of Si evaluated as a function of the radius of sphere B which is the internal sphere in Fig. \ref{spheres1}. Dependence of the results on this parameter appears only because of the differences in basis set sizes between the interior of sphere B and the region outside of sphere B but inside of sphere C. Convergence is achieved when the basis set outside of sphere B becomes good enough so that its gradual replacement with the basis set from the interior of sphere B (when the radius of sphere B increases) does not change the results.

\begin{figure}[t]
\centering
\includegraphics[width=9.0 cm]{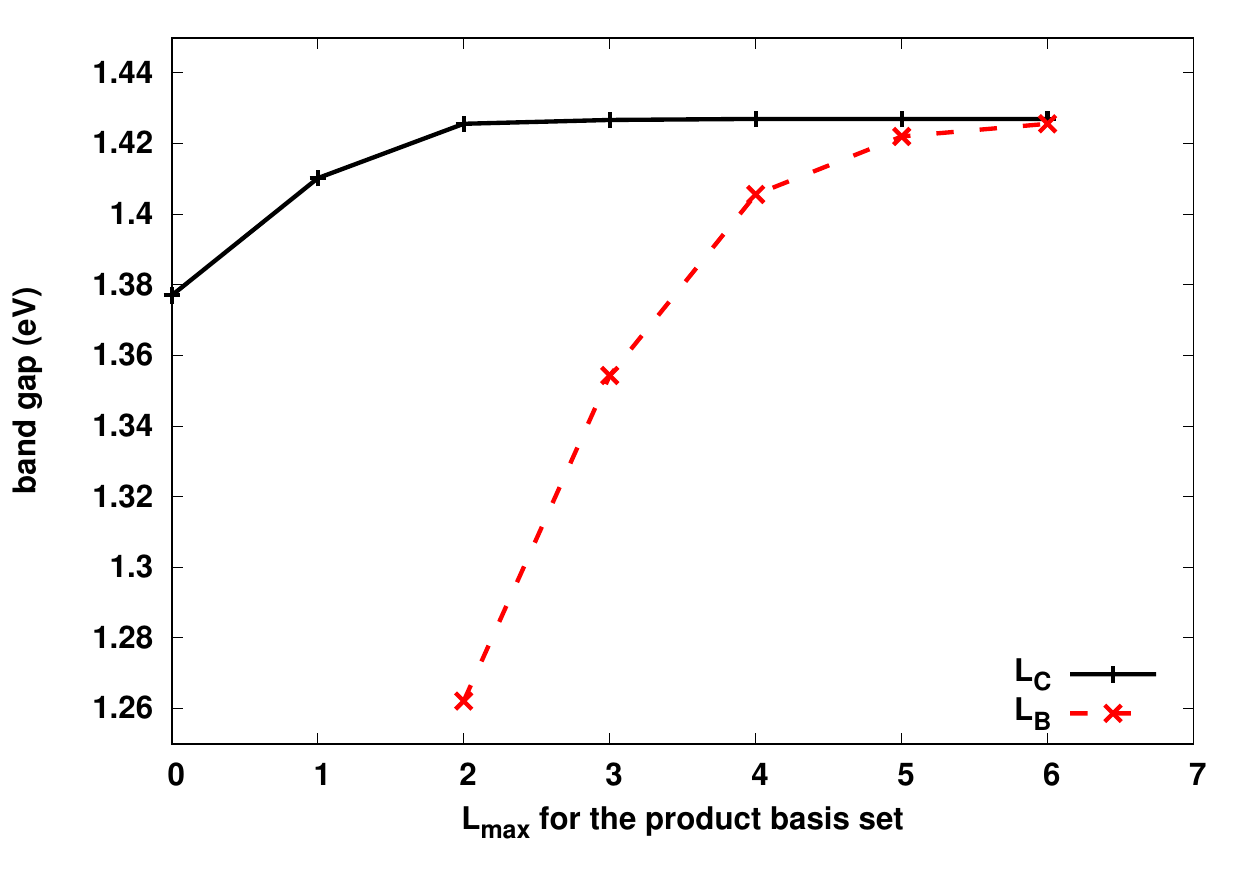}
\caption{Convergence of the calculated band gap of Si with respect to the cutoff $L_{max}$ for the product basis set when the distance between two spheres is smaller than the limiting radius B and when the distance is larger than radius B but smaller than the radius C. For the first case, $L_{max}$ for all C-pairs was fixed at 2, for the second case the $L_{max}$ for the B-pairs was fixed at 6.} \label{conv_lpb}
\end{figure}

As one can see from Figure \ref{conv_rbc}, convergence in the case of Si is achieved when the radius of sphere B reaches value 12 a.u. Even when its value is equal to 8 a.u., the calculated band gap is very close to the final answer. Thus, Fig.\ref{conv_rbc} shows the possibility to use reduced basis sets for two-point quantities when their space arguments are sufficiently far away from each other. It is interesting also to look at this from slightly different perspective. Namely, if we fix the radii of all spheres and just change the size of the basis sets in different regions of space we will get direct measure of how big those basis sets should be. It would also be another support of the possibility to use compact basis sets for distant cells. Figure \ref{conv_lpb} represents such exploration. It shows two lines - solid and dashed. The solid line represents the band gap of Si as a function of $L_{C}$ (cutoff in the product basis set for the muffin tin spheres in the distant unit cells) with $L_{B}$ (cutoff in the product basis set for the muffin tin spheres in the neighboring unit cells) fixed at $L_{B}=6$. The dashed line shows the band gap of Si as a function of $L_{B}$ with $L_{C}$ fixed at $L_{C}=2$. The basis set in the central sphere (sphere A in the Fig. \ref{spheres1}) was fixed at $L_{A}=6$. The radii of spheres B and C were kept unchanged at $R_{B}=8 a.u.$ and $R_{C}=22 a.u.$. First, it is clear that convergence of the calculated band gap is very fast when one varies $L_{C}$ and its value $L_{C}=2$ is sufficient. Convergence of the calculated gap with respect to $L_{B}$ is slower and is only reached at $L_{B}=6$ which is equal to the fixed value of $L_{A}$. This fact simply indicates that for Si, the physics associated with neighboring atoms is essential. Most important for our study, however, is the fact that the basis set for remoted unit cells can be considerably reduced. The amount of computer memory needed to store two-point quantities is proportional to the number of basis set functions per one pair of atoms and to the number of pairs of atoms. The number of pairs belonging to the region between sphere B and sphere C is usually much larger than the number of pairs belonging to the interior of sphere B as one can conclude from the corresponding ratio $\frac{R_{C}^{3}-R_{B}^{3}}{R_{B}^{3}}$. For our selected geometry ($R_{B}=8 a.u.$, $R_{C}=22 a.u.$) this ration is approximately 20. The storage related to the size of the product basis set is roughly proportional to $(L_{(B)(C)}+1)^{4}$ per one pair which simply tells us that the reduction in memory for remoted cells is $(7/3)^{4}\approx 30$. Combined with the fact that the distant cells represent vast majority of all included cells, the advantages of the new implementation are obvious.

\section{Scalability tests and timings}
\label{scal}

\begin{figure}[t]
\centering
\includegraphics[width=9.0 cm]{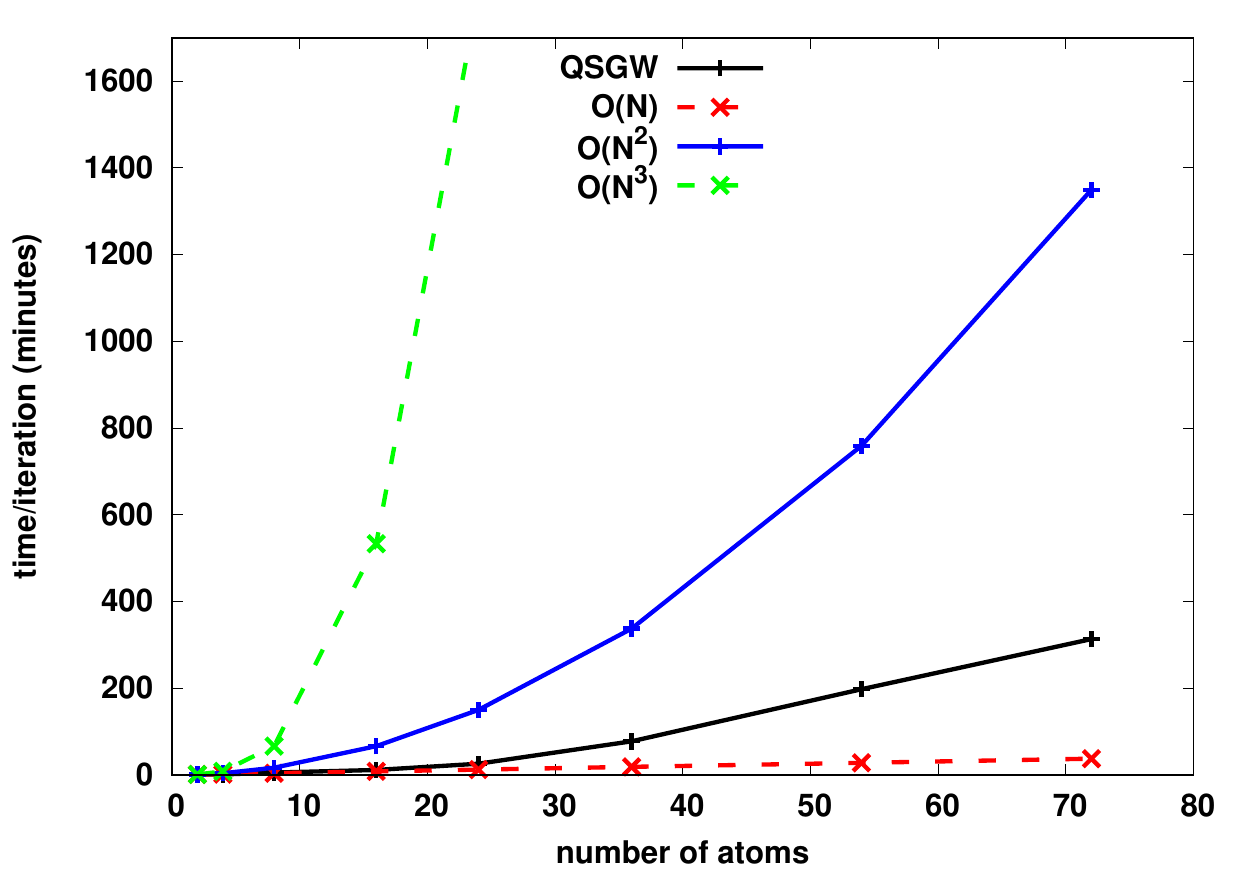}
\caption{Computation time as a function of the number of atoms in the Si supercell. Hypothetical curves with extrapolated $O(N)$, $O(N^{2})$, and $O(N^{3})$ scaling are shown for guidance (see text for the details). Symmetry was switched off in the QSGW runs.} \label{time_size}
\end{figure}

The new implementation of the scGW/QSGW approach combines linear scaling parts (polarizability and self energy) and cubic scaling parts (evaluation of W and G as well as transforms between $\mathbf{R}$ and $\mathbf{k}$ spaces). It is interesting how it scales as a whole. For instance, how the time per full iteration depends on the number of atoms in the unit cell? Figure \ref{time_size} provides this measurement performed on the supercells of silicon atoms. The sizes of the supercells are varied from 2 atoms (one unit cell) to 72 atoms ($4\times 3\times 3$ supercell). In all cases the same number of computer cores (288) was used. In order to facilitate the analysis, the coefficients $\alpha,\beta,\gamma$ in assumed linear, quadratic, and cubic scaling (i.e. assuming that the time scales as $\alpha N$, or as $\beta N^{2}$, or as $\gamma N^{3}$) were found based on the measured time for 2-atom unit cell. Corresponding lines have been added to the graph. In order to make the test more transparent, the symmetry was switched off. Quick observation from Fig. \ref{time_size} is that up to about 20 atoms the scaling is perfectly linear. Even when the size of the supercell becomes 72 atoms the scaling is still between linear and quadratic, demonstrating the efficiency of the algorithm. A couple of comments are, however, needed. First, as it was explained above, the increase in the size of the supercell automatically results in the reduction of the number of $\mathbf{k}$-points in the Brillouin zone (as $1/N$) until only one $\mathbf{k}$-point is left. The results on the graph were obtained with the number of $\mathbf{k}$-points in the range from 125 (2 atoms) to 18 (72 atoms). In other words, even for 72-atom supercell the limit of just one $\mathbf{k}$-point has not been reached yet. The cubic scaling parts of the algorithm always come with the prefactor $O(N_{\mathbf{k}})$ and the $1/N$-type reduction in the number $N_{\mathbf{k}}$ makes them scale quadratically until the limit of only one $\mathbf{k}$-point is reached. Another comment is related to the parallelization strategy which was optimized for the large supercells whereas for small supercells it was accepted to be the same (in order to make sure that only the number of atoms changes). Consequently, the parallelization strategy for the small supercells was not necessarily optimal. The optimization would, probably, push the QSGW curve on the graph towards the quadratic scaling a bit. However, it will remain lower than pure quadratic scaling because the ratio of the two last points on QSGW curve still is smaller than the corresponding ratio on the pure $O(N^{2})$ curve.

\begin{figure}[t]
\centering
\includegraphics[width=9.0 cm]{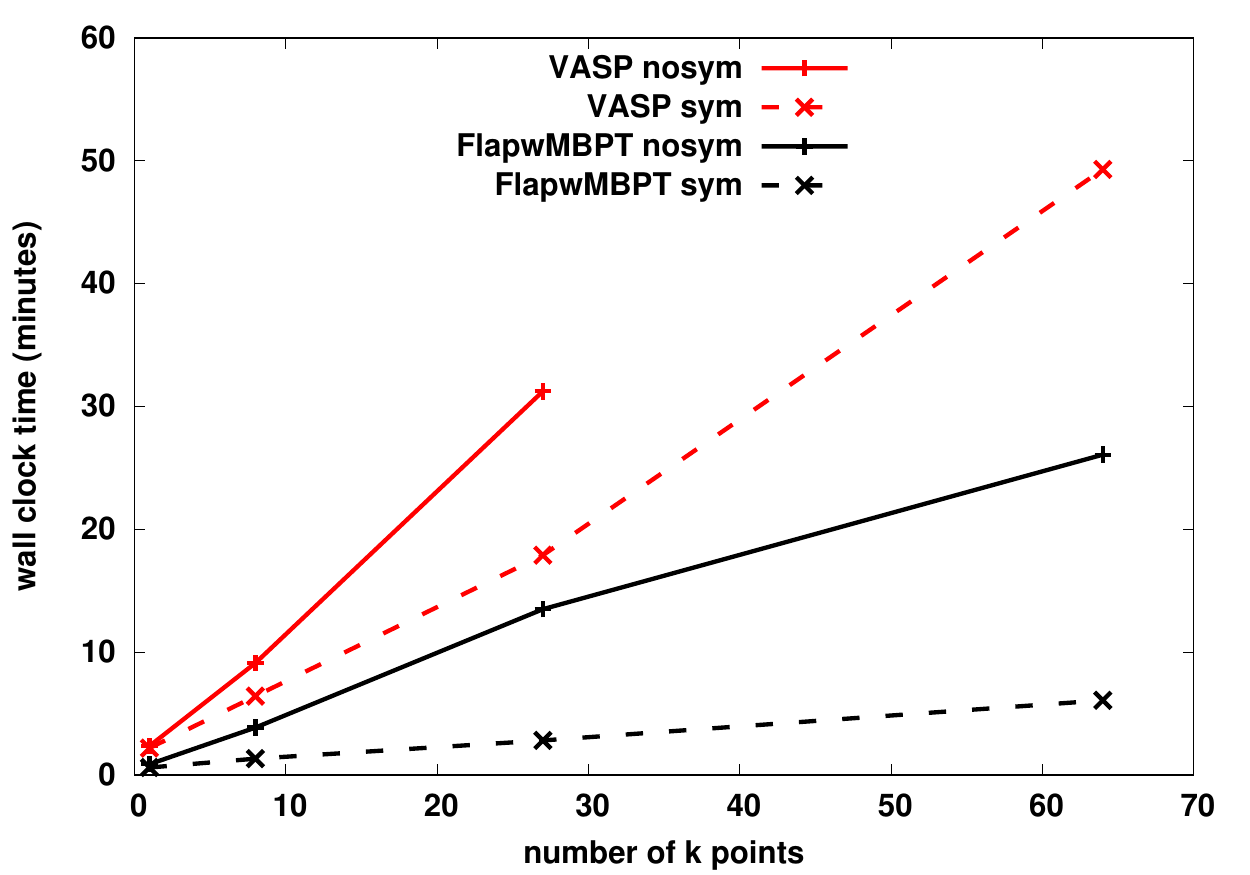}
\caption{Computation time as a function of the number of $\mathbf{k}$-points in the Brillouin zone. 16 Si atom supercell and 64 cores are used. Solid lines represent calculations with the symmetry switched off, whereas the dashed lines represent the calculations performed with full symmetry use. FlapwMBPT timings are given in black color. VASP timings are also given for comparison (in red color).} \label{time_k}
\end{figure}

It is interesting to know, at least approximately, how the performance of the new algorithm compares to the performances of other known codes. Figure \ref{time_k} provides such comparison. As the authors of the work \cite{prb_94_165109} did (using the VASP code \cite{vasp}), the 16-atom supercell of Si atoms was used in the QSGW calculations with 64 cores. One can see, that for this case, the FlapwMBPT code is a few times faster than the VASP code especially for larger number of \textbf{k}-points. A few comments, however, are to be said about the comparison. Certain (essential) difference comes from the fact, that in \cite{prb_94_165109} the time corresponds to $G_{0}W_{0}$ calculations, whereas in this work the time per one QSGW iteration was measured. The full QSGW iteration is more costly than the $G_{0}W_{0}$ because of the need to update the chemical potential and to evaluate full Green's function which was expanded in almost 1000 band states for the 16-atom supercell. In the $G_{0}W_{0}$ calculations \cite{prb_94_165109}, the correction was done only for 64 band states. On the other hand, the measured $G_{0}W_{0}$ times (possibly) include the times for the initial LDA iterations. In the QSGW, the time associated with a few initial LDA iterations was not included, but it comprised in every case less than 10\% of the time per one GW iteration and, consequently, would not change the graphs noticeably. The next factor, potentially affecting the comparison is the differences in the type of computer nodes. Whereas this factor is important, it should not have strong impact on the comparison. The speed of single cores did not change much since 2016 (the year when the work \cite{prb_94_165109} was published). Most important impact may come from the differences in the size of the basis set which is attributed to the intrinsic approach-related part (APW+lo+LO versus PAW) and the part related to the degree of convergence. In this work, the sizes of the basis sets (fermionic and bosonic) were selected such that the convergence of the band energies near the Fermi level was at the level of 0.01eV (see Table \ref{tbas}). This resulted in the sizes of about 1400 (size of the APW+lo+LO basis set) and about 3500 (size of the product basis set). Unfortunately, the authors of \cite{prb_94_165109} do not provide information about their basis set sizes used for the time measurements which, of course, complicates the comparison. In order to facilitate possible future (more detailed) comparison, the dependence of the time per one QSGW iteration on the number of the band states used to represent Green's function is provided in Table \ref{tbas}. Table \ref{tbas} also shows the corresponding band gaps which can help to monitor the level of the convergence.

\begin{table}[t!]
\caption{Dependence of the time/iteration in the QSGW calculations (16-atoms Si supercell) on the size of the basis set. As a measure of the size of the basis set we use the number of the band states to represent Green's function ($N_{bnd}$). The size of the APW+lo+LO basis set was fixed at 1390 functions. As the sizes of the basis sets are slightly different depending on the point in the Brillouin zone, all numbers in the Table related to the size of the basis set were taken at the center of the Brillouin zone. The $4\times 4\times 4$ mesh of the \textbf{k}-points in the Brillouin zone was used. The values of the calculated band gap are provided to show the level of convergence.} \label{tbas}
\begin{center}
\begin{tabular}{@{}c c c}$N_{bnd}$&Time (min)& Band Gap (eV)\\
\hline
327 & 3.68 & 1.356\\
400 & 4.03 & 1.391\\
467 & 4.39 & 1.407\\
543 & 4.82 & 1.409\\
616 & 5.05 & 1.407\\
689 & 5.18 & 1.424\\
761 & 5.41 & 1.417\\
825 & 5.65 & 1.424
\end{tabular}
\end{center}
\end{table}

\section{Applications to real materials}
\label{RM}

Application of the new scGW algorithm to real materials did not pursue the goal of performing thorough study of any specific solid. Rather, the goal of this work was to demonstrate a potential of the code as a tool to study sufficiently large systems containing atoms with $d$ or $f$ electrons. Despite the fact that the choice of materials was a bit random, one can hope that the first application of the new scGW methodology to the selected compounds will open a path for more detailed studying of them (and of other materials of similar complexity) in the future.

\subsection{La$_{2}$CuO$_{4}$}
\label{la2cuo4}

\begin{table}[b]
\caption{Band gap (eV) of La$_{2}$CuO$_{4}$ and magnetic moment (Bohr magnetons) at Cu sites evaluated for different supercell sizes using scGW approach. Principal setup parameters are given for references. Experimental band gap is about 2eV \cite{prb_37_7506,prb_42_10785}. Experimental magnetic moment is 0.4-0.8$\mu_{B}$ \cite{prb_52_7334,prb_73_144513,prl_58_2802}.} \label{LCO}
\begin{center}
\begin{tabular}{@{}c c c c}supercell size (atoms) &14 &28&56\\
\hline
computer cores &320 &320  &384\\
APW+lo basis set &1250 &2500  &4920\\
Product Basis set &4950 &9880  &19730\\
$\mathbf{k}$-mesh &$6\times 6\times 3$ &$6\times 6\times 2$  &$4\times 4\times 3$\\
time per iteration (min) &5 &29  &137\\
Magnetic moment (Cu) &0.81 &0.80  &0.82\\
Band gap &1.91 &1.88  &1.96
\end{tabular}
\end{center}
\end{table}

Principal goal of this study was to explore the ability of the new version of the FlapwMBPT code to run sufficiently large unit cells of more complicated materials (as compared to the silicon) with d- or f-electrons. This ability could be useful when one is interested in (for instance) the substitution effects. Calculations for the supercells of La$_{2}$CuO$_{4}$ in the body-centered-tetragonal (bct) structure were performed in the scalar-relativistic approximation assuming the anti-ferromagnetic (AFM) ordering of spin moments. Structural parameters for this study were taken from the Ref. \cite{prb_38_6650}. The AFM ordering dictated that the minimal size of the supercell has two atoms of Cu in it, i.e. 14 atoms total. This minimal supercell was used as a building block for constructing the 28-atom supercell (by doubling the original supercell in Z direction) and the 56-atom supercell (by doubling the original supercell in X and Y directions). The setup parameters were selected based on a few LDA runs for the minimal 14-atom supercell. The most important setup parameters as well as the main results (magnetic moments and band gaps) are collected in Table \ref{LCO}. The number of computer cores as well as the time per one iteration are also given in Table \ref{LCO}. As the scGW calculations for such large systems are sufficiently time consuming (even with the new algorithm), scGW runs were performed only once without studying the convergence with respect to the parameters specific for the GW part (such as the product basis set size). Nevertheless, small differences in the results obtained with the supercells of different size may serve as an evidence of sufficiently good convergence. An essential result of this study is that now it is possible to use self consistent GW approach to study the electronic structure of complex materials with about 50 atoms (and may be more) per unit cell. Particularly for the La$_{2}$CuO$_{4}$ compound, the ability to study supercells opens a path to study the effects of a substitution of La atoms with other elements (such as strontium) which affects the superconducting properties.

\subsection{CoSbS}
\label{cosbs}

\begin{figure}[h]       
    \fbox{\includegraphics[width=3.9 cm]{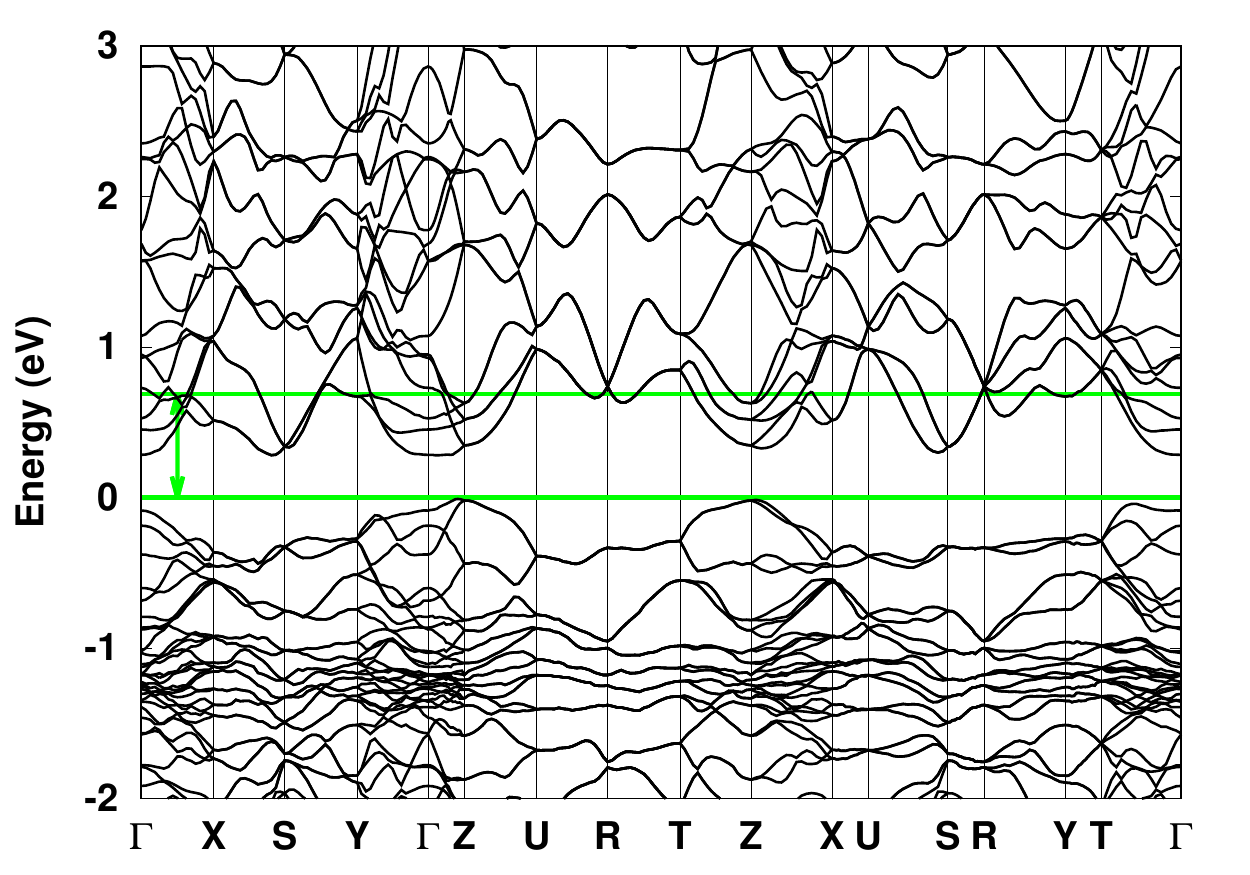}}   
    \hspace{0.02 cm}
    \fbox{\includegraphics[width=3.9 cm]{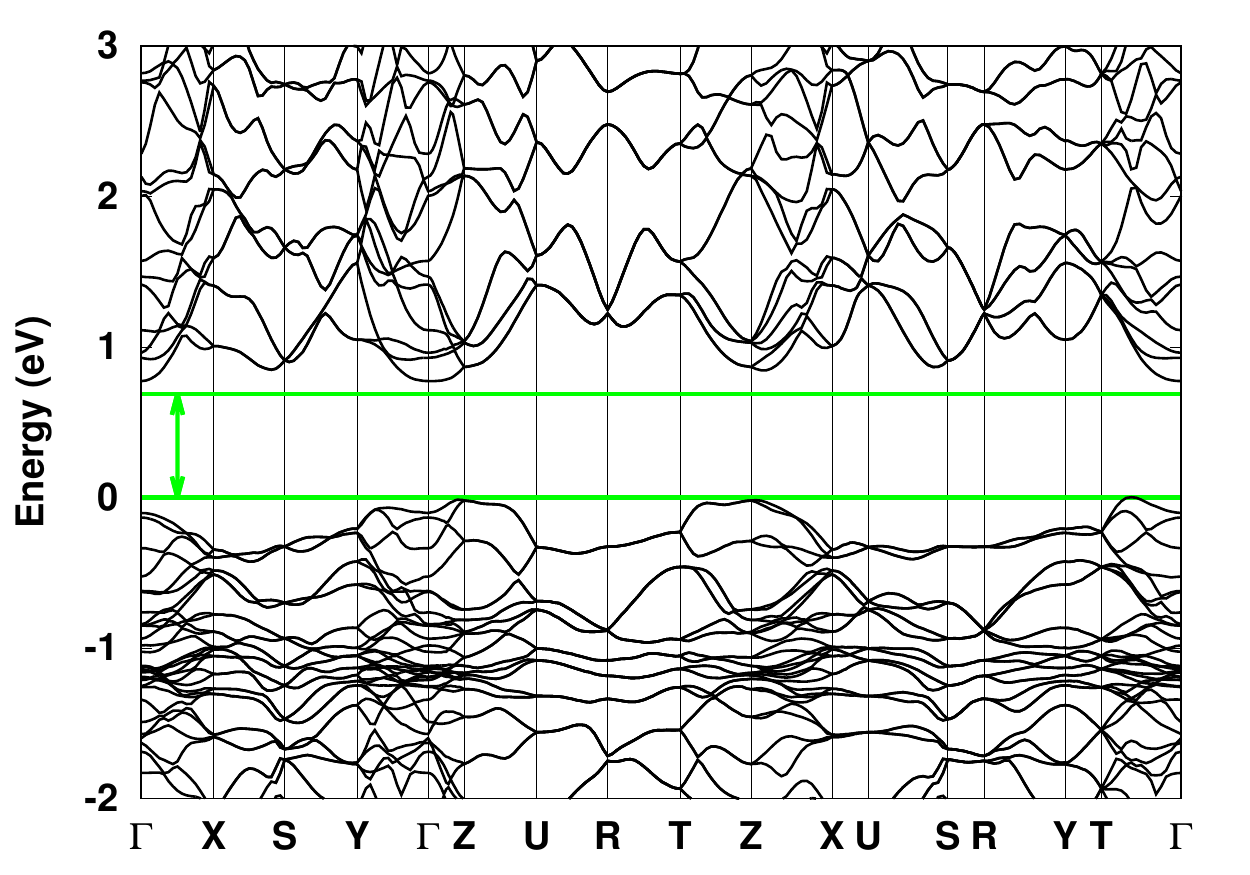}}
    \hspace{0.02 cm}
    \fbox{\includegraphics[width=3.9 cm]{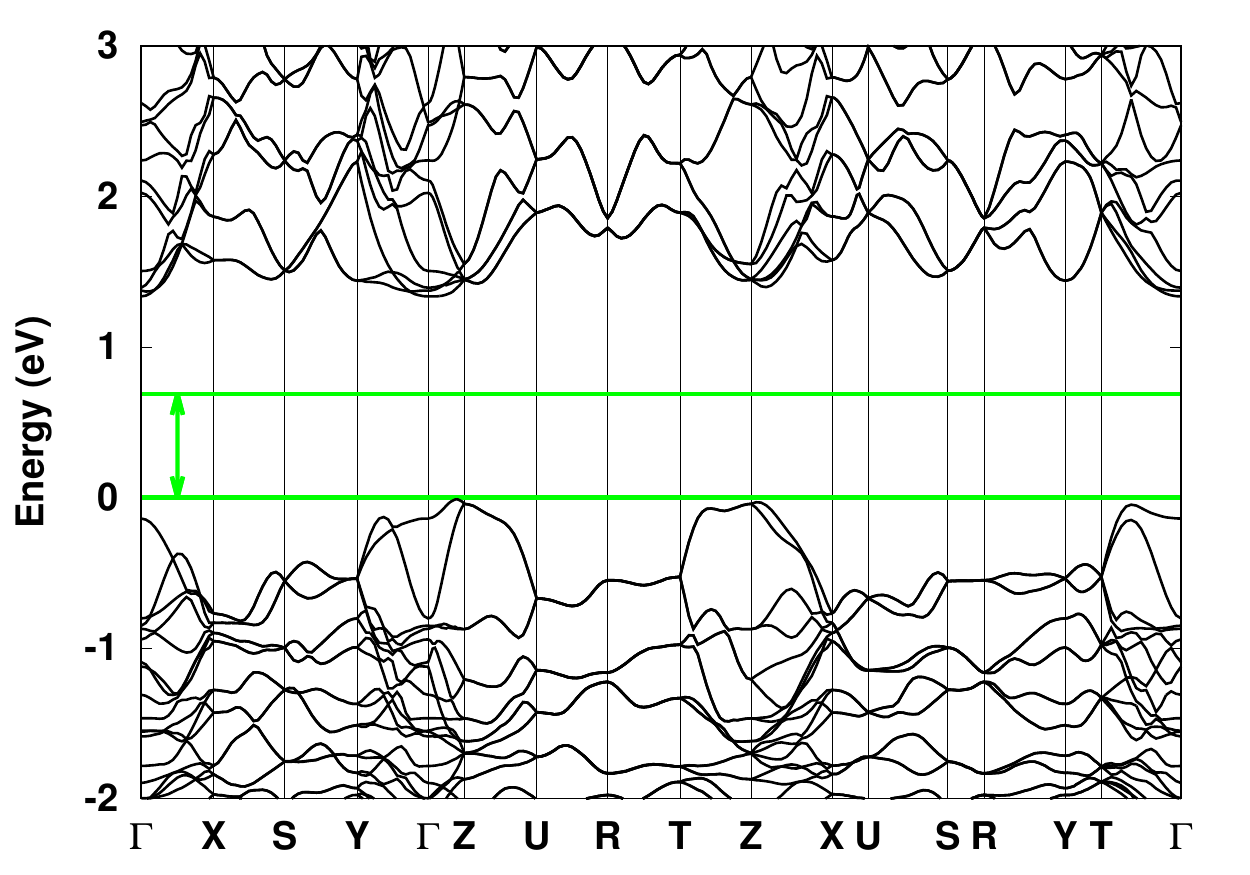}}
    \caption{Electronic band structure of CoSbS evaluated in LDA (left), QSGW (middle), and in scGW (right) approximation.}
    \label{CSS}
\end{figure}

\begin{figure}[h]       
    \fbox{\includegraphics[width=6.0 cm]{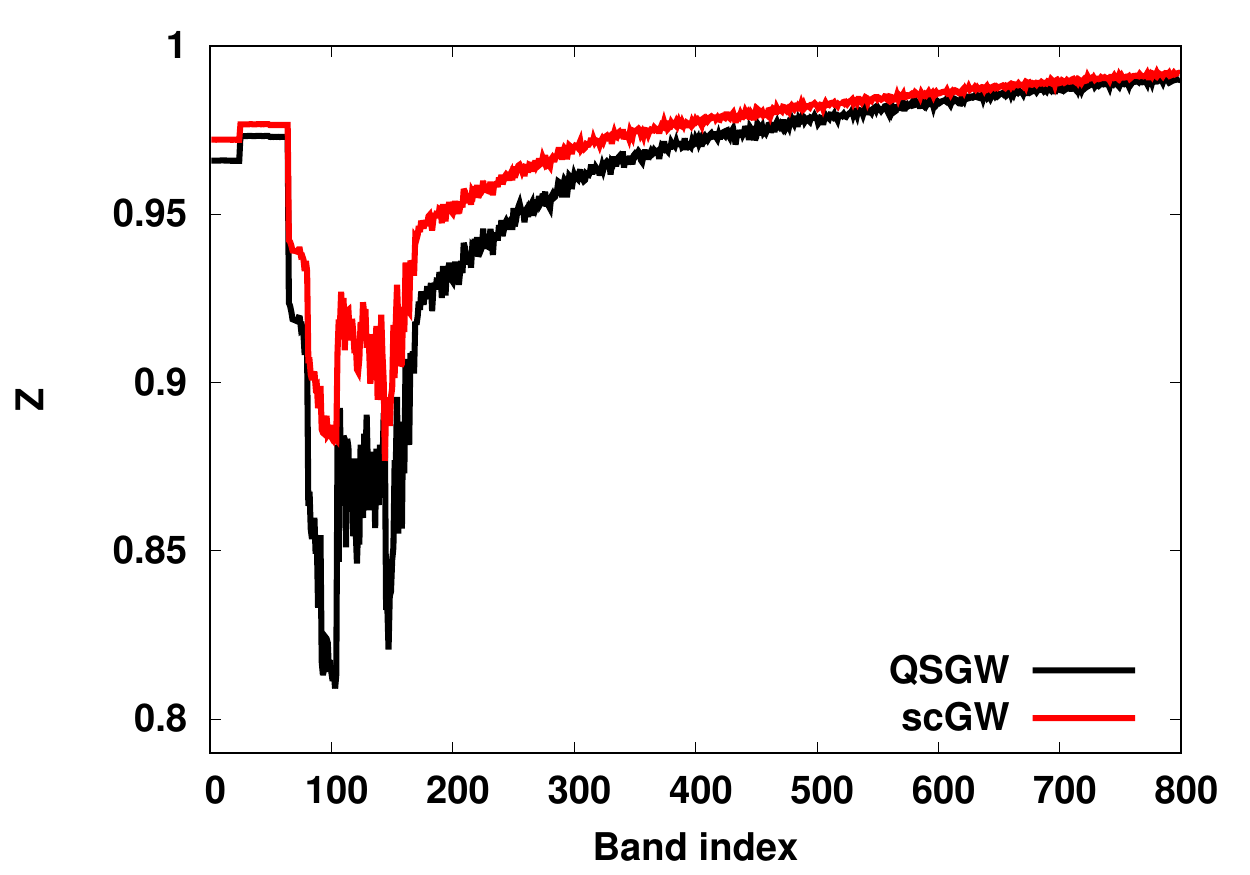}}   
    \hspace{0.02 cm}
    \fbox{\includegraphics[width=6.0 cm]{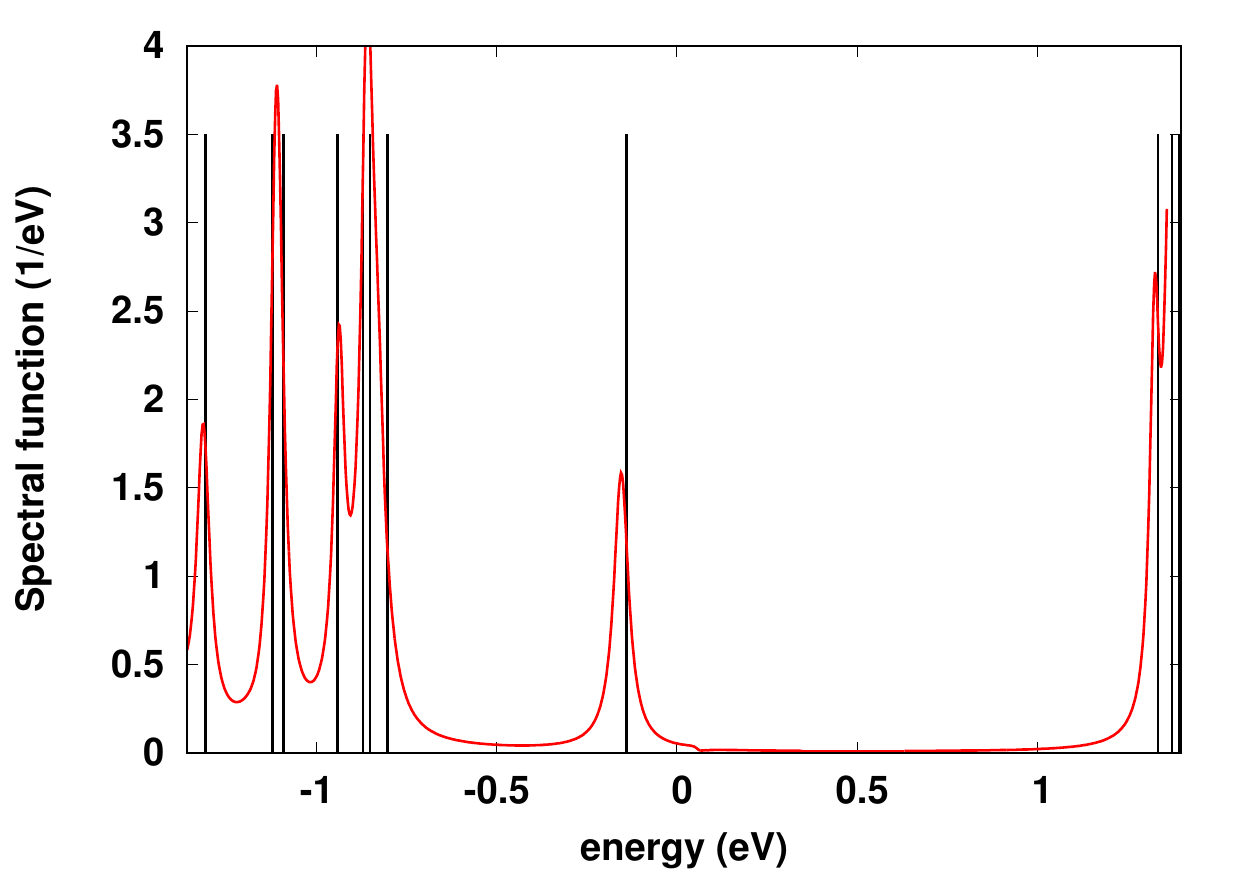}}
    \caption{Left: renormalization Z factor as a function of band index at $\mathbf{k}=0$ obtained in QSGW and scGW approximations. Right: Spectral function (for $\mathbf{k}=0$) of CoSbS obtained in scGW calculation compared with discreet eigen energies from the linearization procedure (see text for the details).}
    \label{ZSF}
\end{figure}

The next compound studied in this work is the paracostibite CoSbS which is an environmetally friendly thermoelectric material \cite{rsca_7_34466}. It crystallizes in the orthorhombic structure (space group 61). The electronic structure of CoSbS was calculated using the three different approximations: LDA, QSGW, and scGW. The experimental structural parameters \cite{rsca_7_34466} were used in all calculations. The numerical setup (cutoffs) for the calculations were selected such that the principal features of the calculated electronic structure (such as band gap) were converged at the level 1-2\% or better. The principal results (band structures) of this study are presented in Figure \ref{CSS}. As one can see, the LDA seriously underestimates the band gap (more than 50\%), whereas the scGW overestimates it (almost by the factor of 2). The best agreement with the experimental value is obtained when one uses the QSGW approximation. Thus, at the moment, the QSGW can be recommended for more detailed studying of this material. However, the QSGW has a disadvantage (similarly to the LDA) of being non-diagrammatic, which represents an obstacle for systematic improvements of these two approaches. In this respect, the fully self consistent GW (which is diagrammatic) represents more fundamental approach and a better platform for improvements. For instance, the diagrammatic vertex corrections \cite{prb_94_155101} which have shown a great success in the simple (primarily consisting from elements with sp-electrons) semiconductors/insulators \cite{prb_95_195120} can also be very useful for CoSbS. The support for this statement comes from the fact that CoSbS is also a weakly correlated material as its renormalization Z-factor (Figure \ref{ZSF}) has values of about 0.87 (and higher) in the scGW and 0.81 (and higher) in the QSGW. These values of the renormalization factor are typical for the weakly correlated semiconductors. An optimization of the algorithm of the vertex corrections represents the next step in the future development of the FlapwMBPT code and it is planned to be accomplished along the same lines as the optimization of the GW part presented in this work.

The plotting of the bands associated with the scGW approach needs to be clarified. Strictly speaking, the one-electron features (band dispersion) in this approach should be obtained as the positions of the peaks of the $\mathbf{k}$-resolved spectral functions. These spectral functions include the analytical continuation of the correlation part of self energy from imaginary to real axis of frequencies. However, as it was demonstrated in the Ref. \cite{prb_85_155129}, the positions of the peaks in spectral function near the chemical potential can often be accurately reproduced by a simplified procedure. This procedure (first introduced in the Ref. \cite{prb_85_155129}) involves the linearization of the frequency dependence of self energy near the chemical potential and, consequently, results in the effective one-electron energies (see details in Appendix \ref{acd}). Figure \ref{ZSF} demonstrates, that this simplified procedure is quite accurate also in the case of CoSbS, at least for the bands surrounding the Fermi energy. The one-electron energies, such obtained, can obviously be used for the plotting purposes.

\subsection{SmB$_{6}$}
\label{smb6}

\begin{figure}[h]       
    \fbox{\includegraphics[width=6.0 cm]{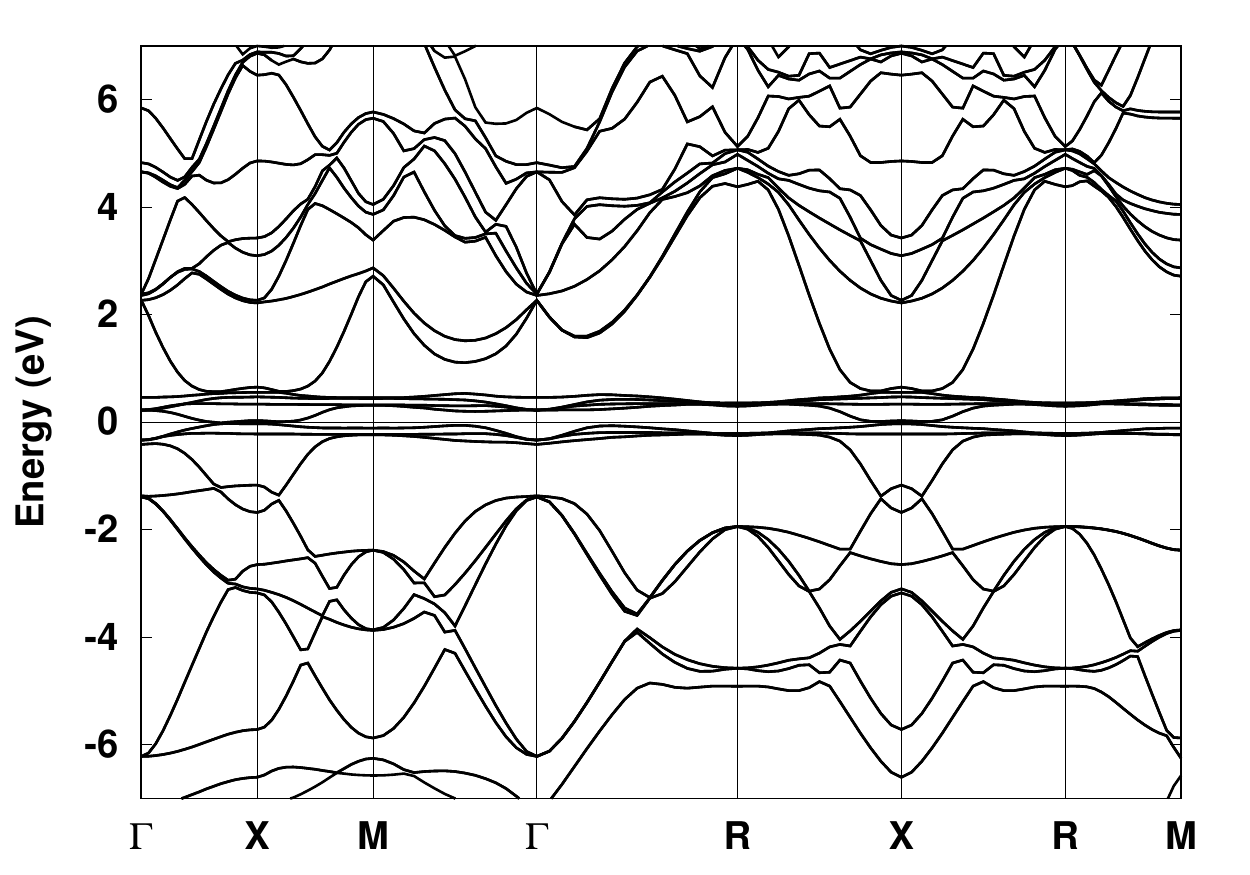}}   
    \hspace{0.02 cm}
    \fbox{\includegraphics[width=6.0 cm]{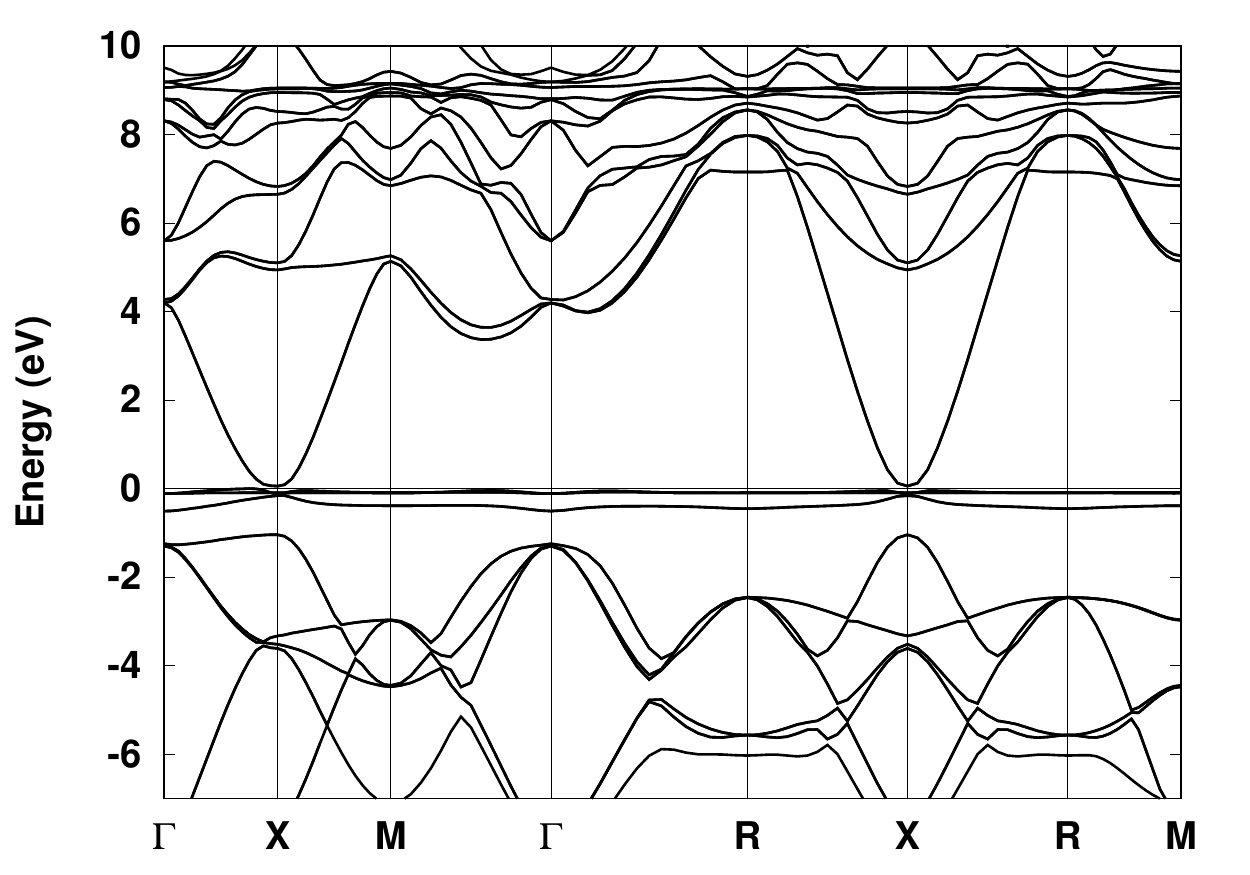}}
    \caption{Electronic band structure of SmB$_{6}$ evaluated in the LDA (left) and in the scGW (right) approximations.}
    \label{SB}
\end{figure}

\begin{figure}[h]       
    \fbox{\includegraphics[width=6.0 cm]{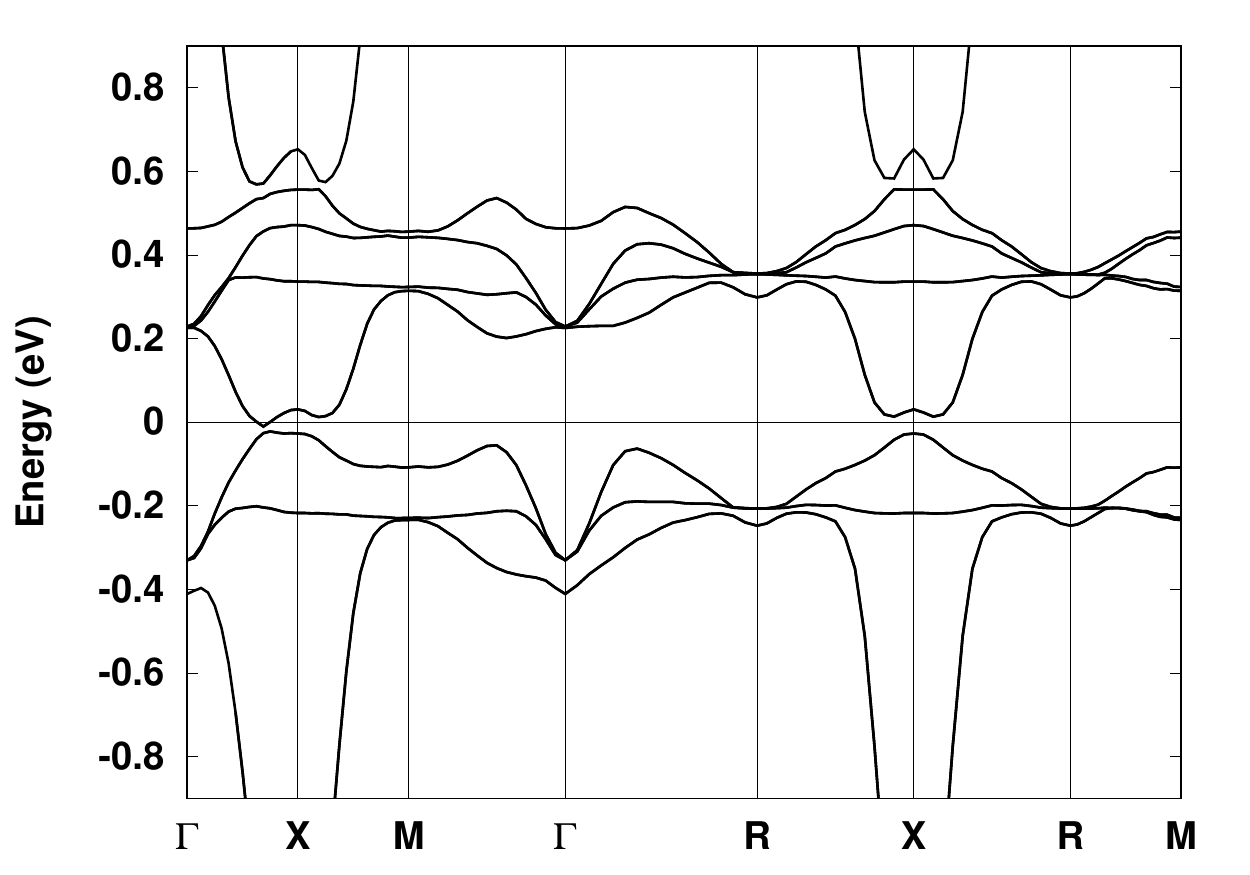}}   
    \hspace{0.02 cm}
    \fbox{\includegraphics[width=6.0 cm]{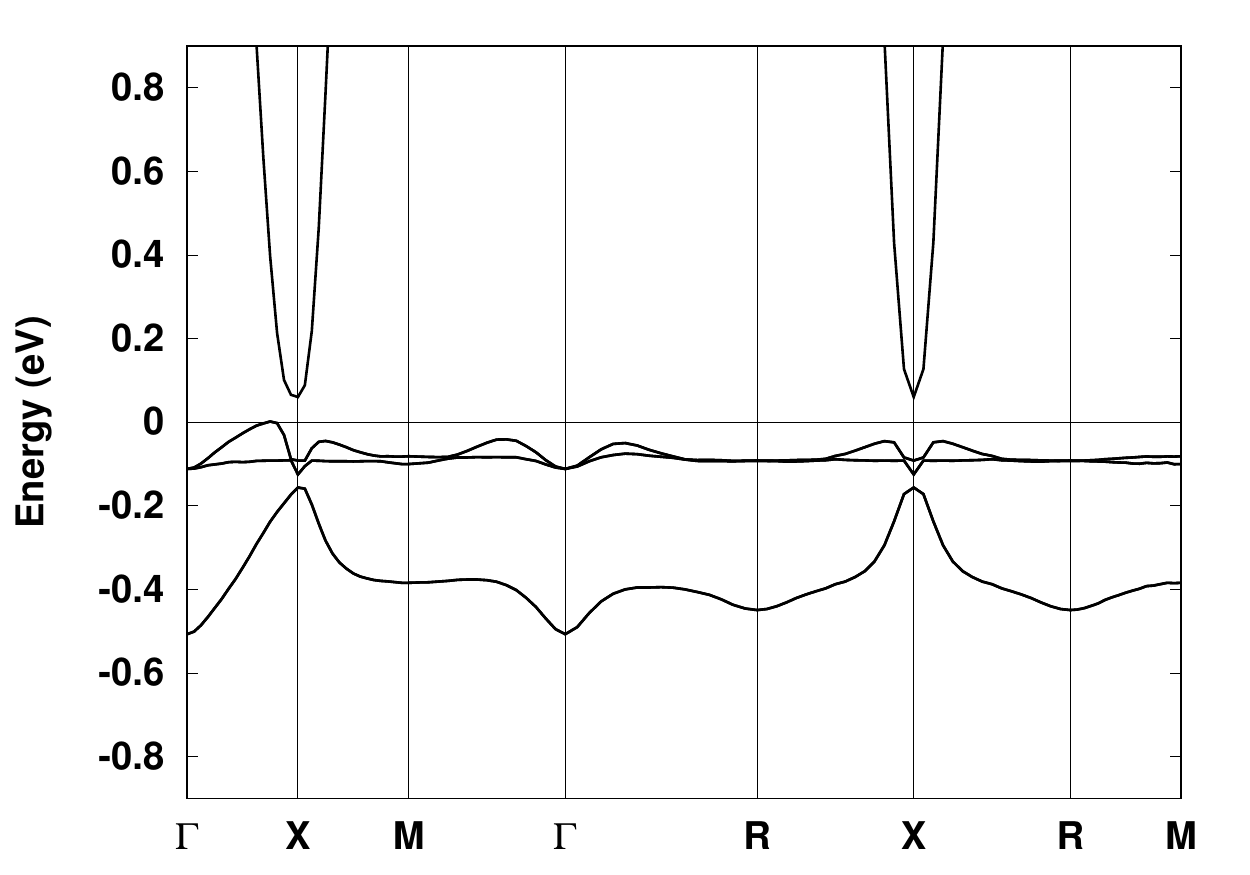}}
    \caption{Low energy electronic band structure of SmB$_{6}$ evaluated in the LDA (left) and in the scGW (right) approximations.}
    \label{SBL}
\end{figure}

\begin{figure}[h]       
    \fbox{\includegraphics[width=6.0 cm]{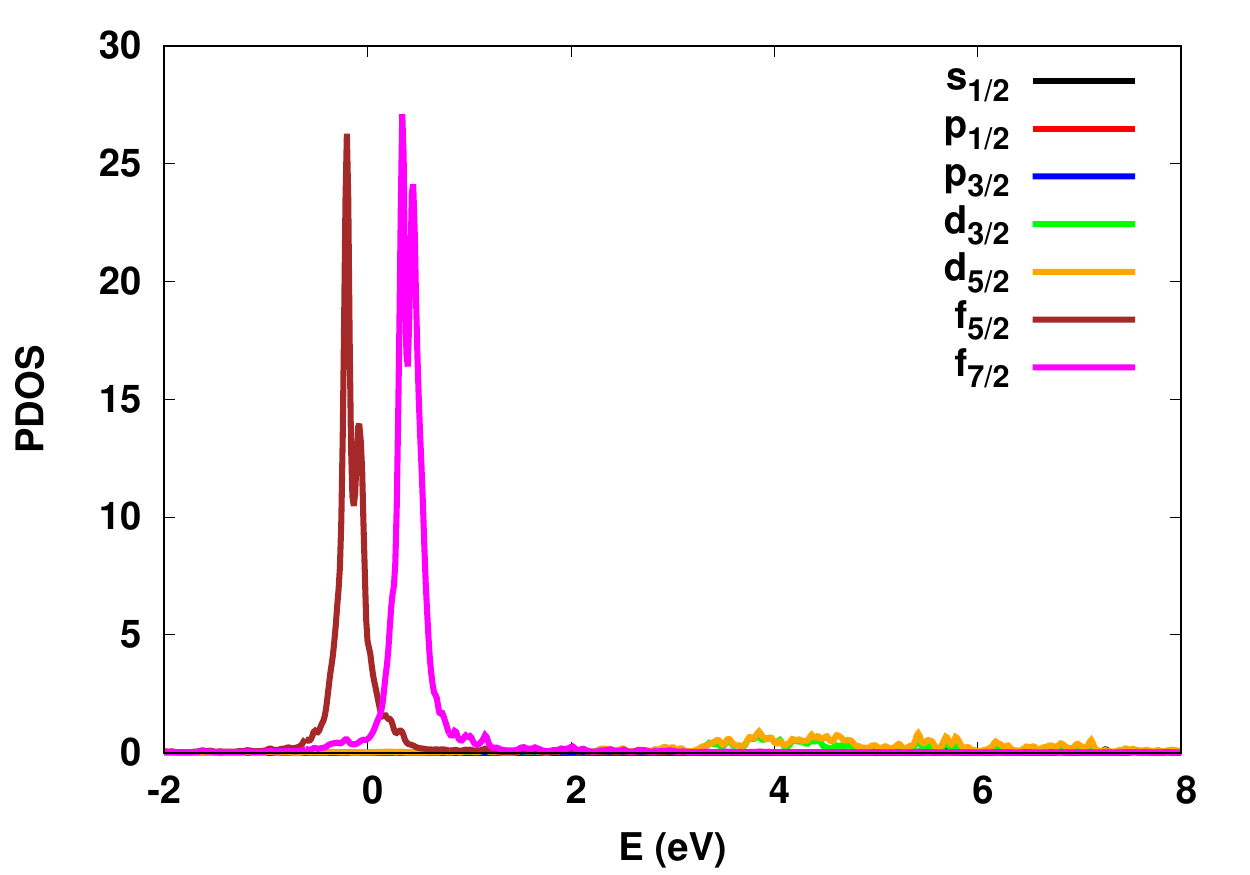}}   
    \hspace{0.02 cm}
    \fbox{\includegraphics[width=6.0 cm]{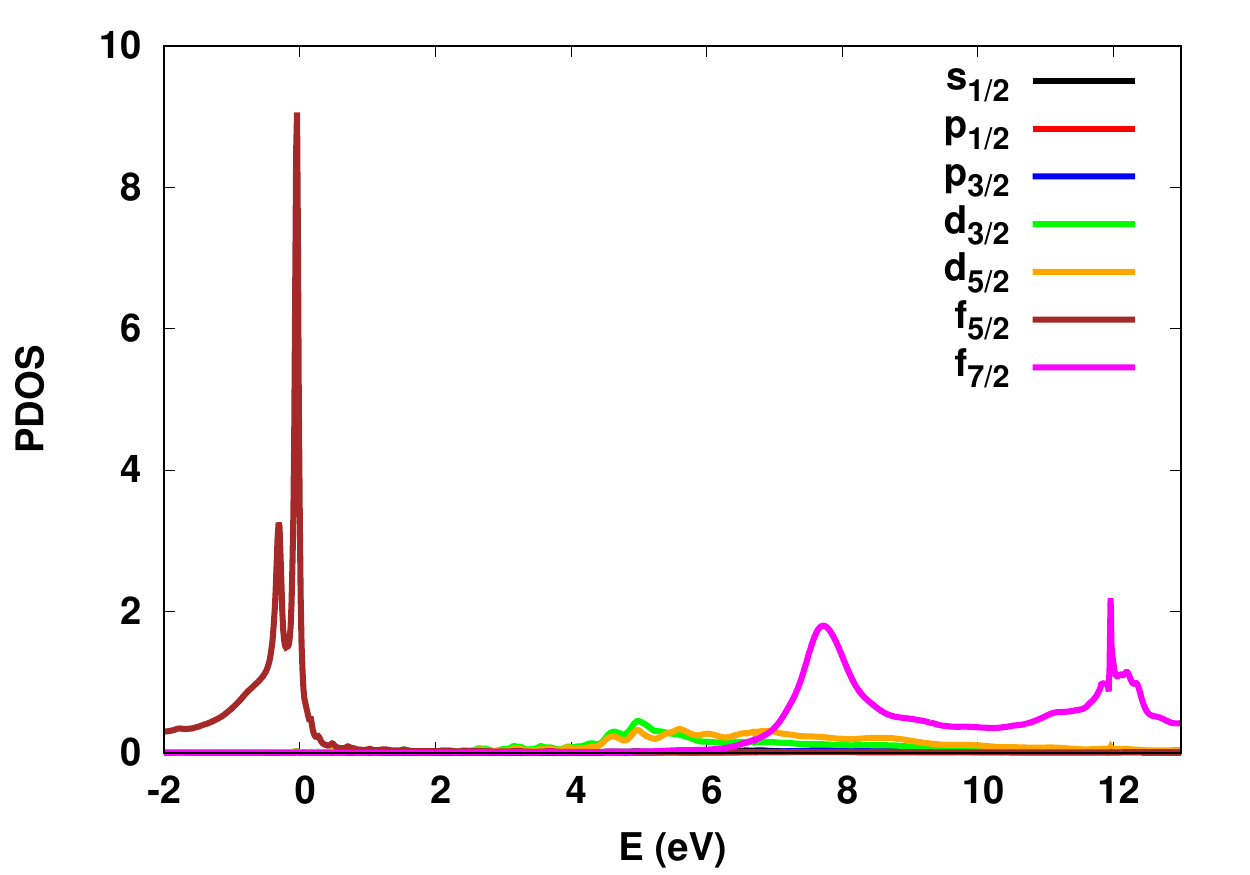}}
    \caption{Partial density of states (spectral function) of SmB$_{6}$ (for Sm atom) evaluated in the LDA (left) and in scGW (right) approximations.}
    \label{PDOS}
\end{figure}

\begin{figure}[h]       
    \fbox{\includegraphics[width=6.0 cm]{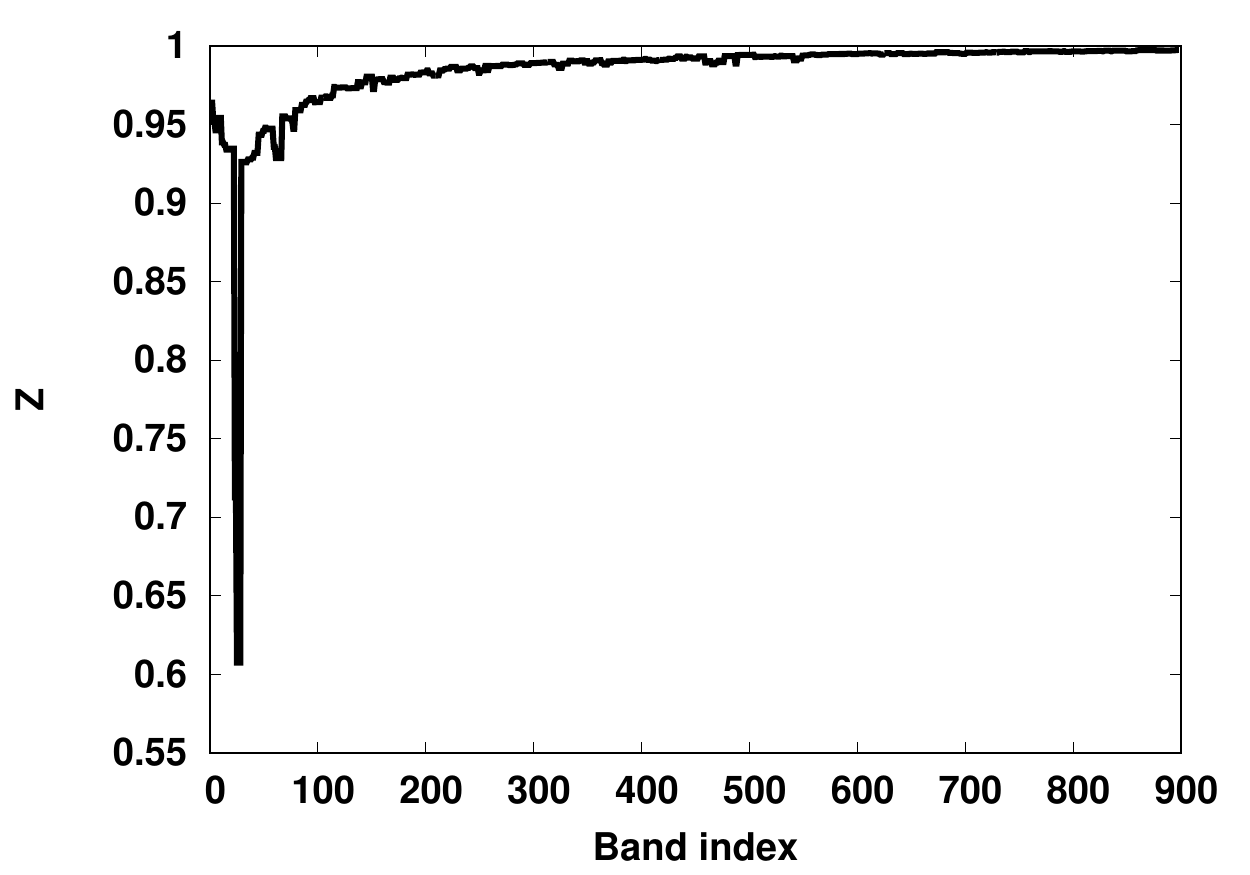}}   
    \hspace{0.02 cm}
    \fbox{\includegraphics[width=6.0 cm]{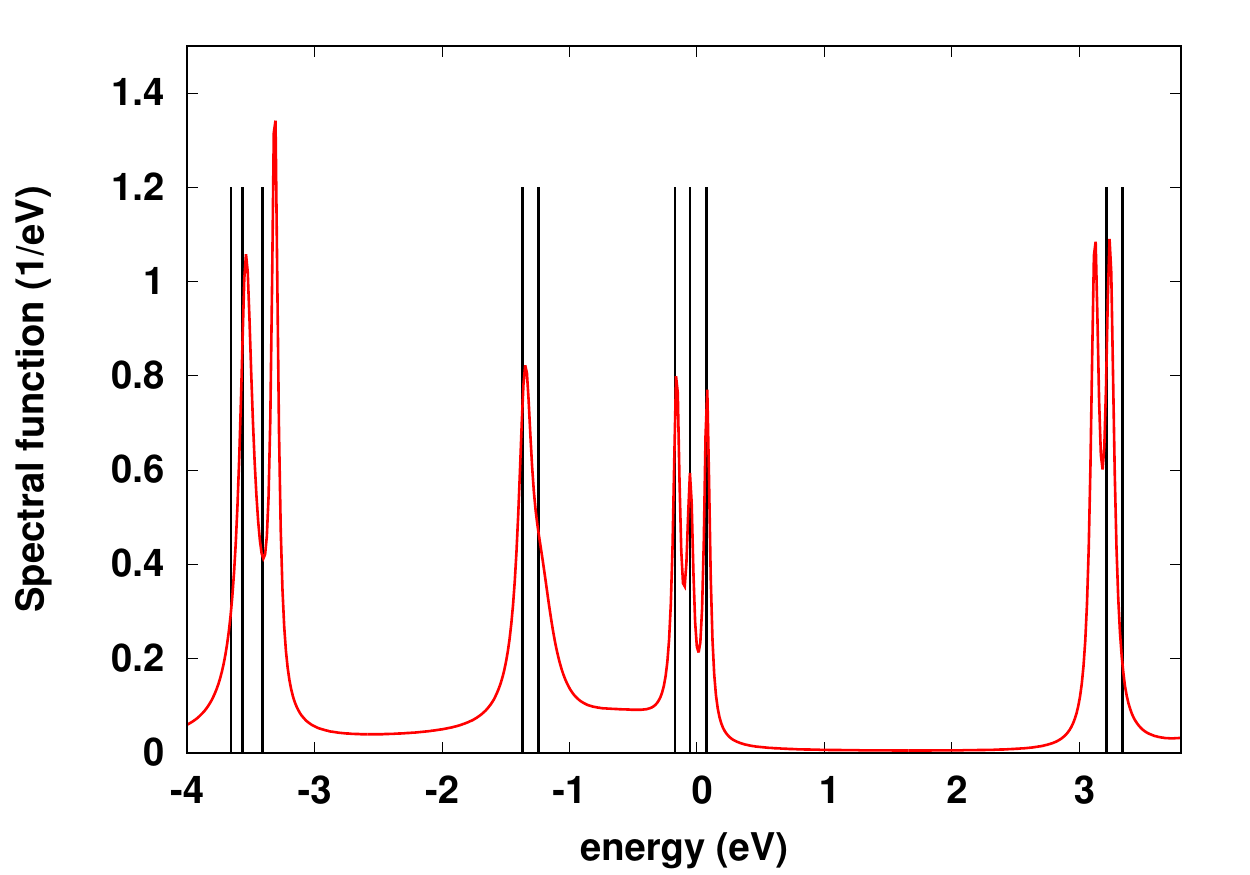}}
    \caption{Left: renormalization Z factor of SmB$_{6}$ as a function of band index at $\mathbf{k}=0$ obtained in the scGW approximations. Right: Spectral function (for the X point in the Brillouin zone) of SmB$_{6}$ obtained in the scGW calculation compared with the discreet eigen energies from the linearization procedure.}
    \label{ZSF_smb6}
\end{figure}

Mixed valence insulator SmB$_{6}$ attracts considerable attention lately in view of its topological properties \cite{prl_110_096401,prb_99_045138,prl_120_016402}. SmB$_{6}$ belongs to the class of the strongly correlated materials and, as such, the GW approximation itself cannot be sufficient for a thorough study of its electronic structure. Previous theoretical studies of this compound were performed using the LDA \cite{ptps_108_19,jpsj_84_024722,prb_66_165209}, the LDA+U \cite{prb_66_165209}, the onsite approximation to the hybrid functional \cite{appa_126_298}, and the LDA+Gutzwiller approach \cite{prl_110_096401}. The LDA approximation is considered as inadequate for this material because it places the f-states near the Fermi level with the splitting between the occupied $f_{5/2}$ and the unoccupied $f_{7/2}$ states of only 0.7eV as compared to the 7eV splitting found in the photoemission expriment \cite{jpfra_41_1141}. The LDA+U approximation misplaces the position of the occupied f-states shifting them to about 5eV below the Fermi level. The LDA+Gutzwiller approximation is designed to take into account the strong electronic correlations (though on the model level) but, similar to the LDA or the LDA+U, it neglects by non-local effects - exchange and screening. It increases the splitting between the occupied and the unoccupied f-states to 4eV, which still is about twice less than in the experiment. In the onsite approximation to the hybrid functional \cite{appa_126_298}, the exchange is taken into account (but only onsite) but the screening is totally missing. The aforementioned splitting between the occupied and the unoccupied f-states is again about 4eV. Both effects (exchange and screening), however, are included automatically on the ab-initio level in the GW-based approaches. Thus, application of the scGW method to SmB$_{6}$ could serve as an additional insight from a different perspective on the calculated electronic structure of this interesting material. The calculations for SmB$_{6}$ in this work were performed for the experimental structural parameters \cite{acsb_66_292} using the fully relativistic approach to account for strong spin-orbit interaction in this material. Principal calculational setup parameters were the following: $8\times 8\times 8$ $\mathbf{k}$-mesh, about 1500 basis functions in the APW+lo basis set (slightly different numbers depending on the $\mathbf{k}$-point), and about 2500 functions in the product basis set.

Figure \ref{SB} shows the band structure of SmB$_{6}$ evaluated with both the LDA and the scGW approaches. As it was anticipated, the principal difference between the two is that the $f_{7/2}$ bands which were situated right above the Fermi level in the LDA calculations have been pushed up to about 7.5-8 eV above the Fermi level in the scGW calculations (see Figure \ref{PDOS} which shows the calculated PDOS). This effect is similar to the one observed earlier in the LDA+Gutzwiller calculations \cite{prl_110_096401} and in the calculations based on the onsite hybrid functional \cite{appa_126_298} where, however, the splitting was insufficient (4 eV). From this comparison one can conclude that the non-local effects included in the scGW are important. Small (0.5-1 eV) overestimation of the splitting in the scGW calculation should obviously be corrected when the vertex corrections are taken into account. The low energy part of the calculated band structure (Figure \ref{SBL}) shows some similarities of the scGW results with the LDA+Gutzwiller band structure obtained in the Ref.\cite{prl_110_096401}. There are differences as well, the principal one being the wider range (beginning at about -0.4eV) of the $f_{5/2}$ bands in the scGW whereas in the LDA+Gutzwiller approach the $f_{5/2}$ bands are in the narrow window (less than 0.1 eV wide) under the Fermi level. Obviously, adding the diagrams to the scGW should shrink the bands but to what degree is hard to estimate without actual calculations. On the other hand, the experimental photoemission result \cite{physb_281_716} indicates the presence of the strong $^{6}H$ multiplets in the range from -0.2 eV to almost the Fermi level which could be more consistent with the scGW results, especially if one takes into account the probable "shrinkage" of the bands associated with the vertex corrections. 

More serious insufficiency of the scGW approximation is obvious from the fact, that the experimental flat band representing the $^{6}F$ multiplet \cite{prb_86_075105} at about 1 eV below the Fermi level is totally missing. It is interesting, however, that in the LDA+Gurzwiller \cite{prl_110_096401} approximation this flat band is also missing. Thus, it looks like the proper on-site physics (supposedly being captured in the Gutzwiller approximation) and the non-local effects (captured in the scGW) should be combined together in order to properly address the physics of SmB$_{6}$.

Figure \ref{ZSF_smb6} (left) demonstrates that SmB$_{6}$ has the renormalization Z-factor of about 0.6 (lowest value) at the scGW level. This value will naturally be reduced when the higher order vertex corrections are included, but even at the scGW level one can see the difference in the strength of the correlations in this material as compared to, for instance, CoSbS (Fig. \ref{ZSF}) where the lowest value of the renormalization factor is about 0.87 (in scGW).

Figure \ref{ZSF_smb6} (right side) demonstrates that also in the case of SmB$_{6}$ the linearization procedure used to plot the bands obtained in the scGW runs is quite accurate. As one can see, all features on the proper spectral function (i.e. the one obtained after the analytical continuation of self energy) are very accurately reproduced by the linearization procedure.

\section*{Conclusions}
\label{concl}

New optimized algorithms for polarizability and self energy in the scGW/QSGW method for solids have been developed. Their implementation in the code FlapwMBPT shows linear scaling with respect to the number of atoms in the unit cell and allows flexible choice of the size of the basis set. Tests and applications to the real materials demonstrate the efficiency of the new approach and its usefulness in material science.

\section*{Acknowledgments}
\label{acknow}
This work was   supported by the U.S. Department of energy, Office of Science, Basic
Energy Sciences as a part of the Computational Materials Science Program.

\appendix

\section{Analytical continuation details}
\label{acd}

The standard definition of spectral function (resolved on the mesh of $\mathbf{k}$-points in the Brillouin zone) is given by

\begin{align}\label{sf_1}
A^{\mathbf{k}}(\omega)=-\frac{1}{\pi}\sum_{\lambda}\mathrm{Im}
G^{R,\mathbf{k}}_{\lambda\lambda}(\omega),
\end{align}
where $\omega$ is a real frequency, $\lambda$ represents band states, and $G^{R}$ is retarded Green's
function. In order to get spectral function in the fully self-consisted GW calculation we analytically continue the correlation part of self--energy $\Sigma _{\lambda \lambda ^{\prime
}}^{c}(\mathbf{k};\omega )$ in order to reconstruct retarded Green's
function on the real axis

\begin{align}\label{sf_7}
G^{^{R},-1}_{\lambda\lambda'}(\mathbf{k};\omega)=(\omega+i\delta+\mu-\varepsilon^{\mathbf{k}}_{\lambda})\delta_{\lambda\lambda'}
-\Sigma^{^{R},c}_{\lambda\lambda'}(\mathbf{k};\omega).
\end{align}

In the formula (\ref{sf_7}), $\delta$ represents a small shift in upper complex plane of frequencies to avoid singularities, $\mu$ is the chemical potential, and $\varepsilon^{\mathbf{k}}_{\lambda}$ are one electron energies from an effective exchange problem \cite{prb_85_155129}, i.e. the eigen-values of the hamiltonian with the correlation part of the self-energy (frequency dependent part) being neglected (the Hartree-Fock like approach).

The analytical continuation is based on the continued fraction expansion
\cite{jltp_29_179} where the self--energy is approximated by the following
expression (for each $\mathbf{k}$ point and for each pair ($\lambda\lambda'$) of bands)

\begin{align}\label{sf_9}
\Sigma(\omega)=\frac{a_{0}}{1+}\frac{a_{1}(\omega-i\omega_{0})}{1+}...\frac{a_{M}(\omega-i\omega_{M})}{1},
\end{align}.

The coefficients $a_{n}$ are found by recurrent relations based on the
values of self--energy at $M+1$ imaginary frequencies. In older implementation of the FlapwMBPT \cite{prb_85_155129} we used 40--80 lowest Matsubara's frequencies (equidistant with step $\frac{2\pi}{T}$).
Whereas the results were quite stable, it was not an optimal choice as it seems now. The inconvenience was that in the self-consistence loop, the frequency points where the functions are evaluated and stored are not distributed equidistantly. Normally, there are 6-20 equidistant points (lowest frequencies), then 20-30 intermediate points with progressive distance between them (in this region an interpolation is used to transform the frequency dependence into the time-dependence), and finally the contribution from asymptotic region is evaluated analytically. All these grids of frequencies were described in details in \cite{prb_85_155129}. So, specially for the analytical continuation a new set of self-energies had to be evaluated at the equidistant points (by special transform from the function of time). Another inconvenience was that the spectral function beyond about 5-10 eV from the chemical potential was affected by how many equidistant frequency points was used in the continuation.

The new implementation is slightly different. Instead of using about 50 equidistant points as an input to the analytical continuation, the same set of self-energies stored on the non-equidistant grid (the same grid as in the self-consistent process) is used. In other words, the same formula (\ref{sf_9}) is used but with slightly different meaning of the frequencies. This new implementation doesn't require for the special time-to-frequency transform and, in addition, shows better stability and robustness for the real frequencies interval up to 15-20 eV from the chemical potential which is good for usual purposes.

In the QSGW calculations we have access to the quasi--particle
energies $E_{i}^{\mathbf{k}}$. In this case the analytical continuation is avoided, and the $\mathbf{k}$-resolved spectral function is represented by a series of $\delta$-function like peaks at the frequencies equal to the quasi-particle energies. 

The access to the effective quasiparticle energies is provided by the linearization procedure. This linearization (see the details in \cite{prb_85_155129,cpc_219_407}) is based on the value and the derivative (with respect to frequency) of self-energy taken at zero frequency. In the self-consistent QSGW, the so obtained effective quasiparticle eigen-energies $E^{\mathbf{k}}_{i}$ are used to easily construct updated (at the new iteration) Green's function, and, subsequently to use it in the evaluation of polarizability and new self-energy. But the same exactly linearization can be used in the fully self-consistent GW calculations in order to simplify the evaluation of spectral function (only at the postprocessing). Namely, the effective quasiparticle energies can be directly compared with the positions of peaks in spectral function obtained from the full analytical continuation of self-energy. Thus, the linearization of self-energy can also be considered as a very simple variant of the analytical continuation useful to get quick estimate of the position of the peaks of true spectral function. By construction, it is good enough only near the chemical potential. But, as practice shows (see Figs \ref{ZSF} and \ref{ZSF_smb6}), the accuracy (reproduction of the position of the peaks) is quite impressive.



\bibliographystyle{elsarticle-num}

\end{document}